\documentclass[12pt,a4paper]{article}

\usepackage[noblocks]{authblk}
\usepackage{a4wide}
\usepackage{times}

\usepackage{enumitem}
\setlist[itemize]{leftmargin=*}
\setlist[enumerate]{leftmargin=*}
\setlist[description]{leftmargin=*}

\usepackage{siunitx}

\usepackage{amsmath,amssymb,amsbsy,amscd,amstext,mathtools,amsxtra}
\usepackage{amsmath,amssymb,amscd,amstext,mathtools,extarrows}

\newcommand{\PW}{\mathcal{PW}}
\newcommand{\Sp}[1]{\mathrm{#1}}
\newcommand{\dX}{\dot{\XX}}

\newcommand{\Kp}{{K_{p}}}
\newcommand{\Vp}{{\VV_{p}}}
\newcommand{\Np}{{\NN_{p}}}
\newcommand{\Npf}{{\NN_{p}^{f}}}
\newcommand{\Npr}{{\NN_{p}^{r}}}
\newcommand{\Ap}{\AA_{p}}
\newcommand{\Apf}{{\Ap^{f}}}
\newcommand{\Apr}{{\Ap^{r}}}
\newcommand{\Pf}{\PP^{f}}
\renewcommand{\Pr}{\PP^{r}}
\newcommand{\Pp}{\PP_p}
\newcommand{\Ppb}{{\bar{P}_p}}
\newcommand{\Ppf}{{\Pp^{f}}}
\newcommand{\Ppr}{{\Pp^{r}}}

\newcommand{\Na}{\text{Na}^+}
\newcommand{\Nai}{\text{Na}_i^+}

\newcommand{\Nao}{\text{Na}_o^+}

\newcommand{\lb}{\left (}

\newcommand{\Ln}{\text{\bf Ln }}
\newcommand{\Exp}{\text{\bf Exp }}
\newcommand{\rb}{\right )}

\newcommand{\Nf}{N^f}
\newcommand{\Nr}{N^r}

\newcommand{\Ncd}{{N^{cd}}}

\newcommand{\Af}{{A^f}}
\newcommand{\Ar}{{A^r}}

\newcommand{\NN}{N}
\renewcommand{\AA}{{A}}

\newcommand{\PP}{P}
\newcommand{\VV}{{V}}

\newcommand{\XX}{{X}}

\newcommand{\II}{\mathcal{I}}

\newcommand{\kkappa}{\kappa}
\newcommand{\mmu}{{\mu}}

\newcommand{\reacul}[2]{
  {\; \xrightleftharpoons[#2]{#1} \;}
}

\newcommand{\reacu}[1]{
  \reacul{#1}{}
}

\newcommand{\BG}[1]{\text{\sffamily\textbf{#1}}}

\newcommand{\C}{\BG{C }}

\newcommand{\R}{\BG{R }}

\newcommand{\one}{\BG{1 }}
\newcommand{\zero}{\BG{0 }}

\newcommand{\TF}{\BG{TF }}

\renewcommand{\Re}{\BG{Re }}

\newcommand{\BGL}[2]{$\BG{#1}$:$\mathbf{#2}$} %Generic

\newcommand{\BC}[1]{\BGL{C}{#1}}

\newcommand{\BRe}[1]{\BGL{Re}{#1}}

 \usepackage[numbers,compress]{natbib}
\usepackage{graphicx,subfigure,fancybox,color}

\newcommand{\SubFig}[3]{
 \subfigure[#2]{
   \includegraphics[width=#3\linewidth]{#1.pdf}
   \label{subfig:#1}
 }
}

\usepackage{doi}
\usepackage{url}

\begin{document}

% \makeatletter
% \@namedef{Changes@AuthorColor}{magenta}
% \colorlet{Changes@Color}{magenta}
% \makeatother

\title{Energy-based Analysis of Biomolecular Pathways}
\author{Peter J. Gawthrop\footnote{Corresponding author. \textbf{peter.gawthrop@unimelb.edu.au}}}
\affil{
  Systems Biology Laboratory,
  Department of Biomedical Engineering,
  Melbourne School of Engineering,
  University of Melbourne,
  Victoria 3010, Australia.
  \authorcr
  Department of Electrical and Electronic Engineering, 
  Melbourne School of Engineering,
  University of Melbourne,
  Victoria 3010, Australia.
   }

\author{Edmund J. Crampin}
\affil{
   Systems Biology Laboratory,
   Melbourne School of Engineering,
   University of Melbourne,
   Victoria 3010, Australia.
   \authorcr 
   School of Mathematics and Statistics,
   University of Melbourne,
   Victoria 3010, Australia.
   \authorcr 
   School of Medicine,
   University of Melbourne,
   Victoria 3010, Australia.
  \authorcr 
   ARC Centre of Excellence in Convergent Bio-Nano Science,
   Melbourne School of Engineering,
   University of Melbourne,
   Victoria 3010, Australia.
}

%  \author{Peter J. Gawthrop$^{1}$  and Edmund J. Crampin$^{1-4}$}
%  \address{$^{1}$
%    Systems Biology Laboratory,
%    Department of Biomedical Engineering,
%    Melbourne School of Engineering,
%    University of Melbourne,
%    Victoria 3010, Australia\\
%    $^{2}$
%    School of Mathematics and Statistics,
%    University of Melbourne,
%    Victoria 3010, Australia\\
%    $^{3}$
%    School of Medicine,
%    University of Melbourne,
%    Victoria 3010, Australia\\
%    $^{4}$
%    ARC Centre of Excellence in Convergent Bio-Nano Science,
%    Melbourne School of Engineering,
%    University of Melbourne,
%    Victoria 3010, Australia}

% \subject{Mathematical modelling; Biochemistry; Engineering}
% \keywords{Network thermodynamics; biomolecular systems; bond graph; reaction kinetics.}
% \corres{Peter J. Gawthrop\\ \email{peter.gawthrop@unimelb.edu.au}}
\maketitle
 \begin{abstract}
   Decomposition of biomolecular reaction networks into pathways is a
   powerful approach to the analysis of metabolic and signalling
   networks. Current approaches based on analysis of the
   stoichiometric matrix reveal information about steady-state
   mass flows (reaction rates) through the network. In this work we
   show how pathway analysis of biomolecular networks can be extended
   using an energy-based approach to provide information about energy
   flows through the network. This energy-based approach is developed
   using the engineering-inspired bond graph methodology to represent
   biomolecular reaction networks.
   The approach is introduced using glycolysis as an exemplar; and is
   then applied to analyse the efficiency of free energy transduction
   in a biomolecular cycle model of a transporter protein
   (Sodium-Glucose Transport Protein 1, SGLT1).
   The overall aim of our work is to present a framework for modelling
   and analysis of biomolecular reactions and processes which
   considers energy flows and losses as well as mass transport.
     %   \deleted{The stoichiometric analysis of metabolic pathways is combined with
   % the energy-based analysis of biomolecular networks based on bond
   % graphs to give an energy-based interpretation steady-state flow
   % pathways.  In particular, a stoichiometric null-space matrix is not
   % only used to identify steady-state flow pathways but also to
   % determine the corresponding pathway affinities and thence pathway
   % energy dissipation. This approach gives a systematic way to
   % understand energy dissipation and energy efficiency in biomolecular
   % networks.
   % The concepts are illustrated using a standard model of glycolysis,
   % a standard model trans-membrane free-energy transduction and a model
   % of the Sodium-Glucose Transport Protein 1 (SGLT1).}
 \end{abstract}

\newpage
\tableofcontents
\newpage

\section{Introduction}
\label{sec:introduction}
%% Stoichiometric anaysis
The term ``pathway analysis'' is used very broadly in systems biology to
describe several quite distinct approaches to the analysis of
biomolecular networks \citep{KhaSirBut12,SanMinSch13,PapStePri04}
% %
% \deleted{(refs including Khatri et al 2012, Sandefur et al 2013, Papin
%   et al 2004 ).}
% %
and often the definition of a ``pathway'' is somewhat nebulous. In
this work we are concerned with those pathways defined in terms of the
stoichiometric analysis of biomolecular reaction networks
\citep{Cla88,HeiSch96,Pal06,Pal11,KliLieWie11}; in particular, the
null space%
\footnote{The mathematical concept of a null space is given a
  biological interpretation in Appendix \ref{sec:short-intr-syst}.}
of an appropriate stoichiometric matrix is used to identify the
pathways of a biomolecular network.
Two alternative concepts of pathways: \emph{elementary modes} and
\emph{extreme pathways} are compared and contrasted by
\citet{PapStePri04}.
Computational issues are considered by \citet{SchHil94},
\citet{PfeSanNun99}, \citet{SchLetPal00} and \citet{SchHilWooFel02}.
A brief introduction to the relevant concepts appears in
Appendix~\ref{sec:short-intr-syst}.

These approaches have proven to be very useful in determining network
properties and emergent behaviour of biomolecular reaction networks in
terms of the pathway ``building blocks'' of these networks. However,
these approaches are solely focused
% on network
% stoichiometry 
on \emph{mass} flows of biomolecular reaction networks.
However, to date, little attention has been given to the
identification and analysis of pathways in the context of
\emph{energy} flows in biomolecular reaction networks. This paper
extends the pathway concept to include such energy flows using an
engineering-inspired method: the bond graph.

%% Bond Graph  stuff
Like engineering systems, living systems are subject to the laws of
physics in general and the laws of thermodynamics in particular
\citep{Lot22,Hil89,QiaBeaLia03,BeaQia10,MarSouLan14}.  This fact gives
the opportunity of applying engineering science to the modelling,
analysis and understanding of living systems.
The bond graph method of \citet{Pay61} is one such energy-based
engineering approach
\citep{KarMarRos12,Cel91,GawSmi96,GawBev07,Bor11,Bor17} which has been
extended to include chemical systems \citep{Cel91,GreCel12},
biological systems~\citep{DiaPic11} and biomolecular systems
\cite{OstPerKat71,OstPerKat73,GawCra14,GawCurCra15,GawCra16,Gaw17}.
A brief introduction to the bond graph  approach appears in
Appendix~\ref{sec:short-intr-bond}.

%% External metabolites, source source and sink species, chemostatas
%% etc
%% Energy based analysis
% Energy based analysis \citep{Hil89}, for example using the bond graph
% approach \citep{OstPerKat71,OstPerKat73,GawCra14,GawCra16}, 
Applying the bond graph method to modelling biomolecular pathways moves
the focus from mass flow to energy flow. 
% In particular,
% the stoichiometric matrix can be used not only to analyse mass flows
% but also to analyse chemical affinities and free energy
% transduction. 
%
%% Main message
Hence this paper brings together stoichiometric pathway analysis with energy
based bond graph analysis to identify the pathways of steady-state free energy
transduction in biomolecular networks. 
%
%% Contribution
Although the bond graph approach is well-established in the field of
engineering, and the stoichiometric pathway analysis approach
well-established in the field of biochemical analysis, they have not
hitherto been brought together. As illustrated in the Sodium-Glucose
Transport Protein example of \S~\ref{sec:exampl-sodi-gluc}, this
interdisciplinary synthesis gives new insight into the energetic
behaviour of living systems.

% Following
% \citet{GawCra14,GawCra16}, a bond graph interpretation is given.

\S~\ref{sec:energy-based-react} introduces the bond graph approach to
pathway analysis using glycolysis as an exemplar.
As a preliminary to the energy-based approach,
\S~\ref{sec:stoich-analys-pathw} discusses the stoichiometric approach
to pathway analysis.
\S~\ref{sec:pathway-analysis} combines the energy-based reaction analysis of
\S\ref{sec:energy-based-react} with the stoichiometric pathway
analysis of \S\ref{sec:stoich-analys-pathw} to give an energy-based
pathway analysis of biomolecular systems.
\S~\ref{sec:example:-free-energy} looks at the generic  biomolecular cycle
transporter model of \citet{Hil89} focusing on the energy transduction
aspects; in particular, a two-pathway decomposition provides insight
into free-energy dissipation and efficiency.
\S~\ref{sec:exampl-sodi-gluc} uses this two-pathway decomposition to
look at a particular transporter -- The Sodium-Glucose Transport Protein
1 (SGLT1) -- using parameters drawn from the experimental results of
\citet{EskWriLoo05}.
\S~\ref{sec:conclusion} draws some conclusions and indicates future
research directions.
Appendix~\ref{sec:short-intr-bond} provides a short introduction to those
features of the bond graph approach necessary to understand this
paper and Appendix~\ref{sec:short-intr-syst} similarly introduces
systems biology.
The complete equations describing the examples are given in the
Supplementary Material.
% \added{In the following, firstly we briefly review energy-based analysis of biomolecular reaction networks using bond graphs, in particular introducing chemostats, which is useful concept in pathway analysis. In particular we illustrate the approach using the example of glycolysis. Next we review stoichiometric pathway analysis, specifically calculation of the null-space matrix of the stoichiometric matrix. We then develop an energy-based approach to pathway analysis, using the bond graph formulation, in order to be able to calculate both molar flows and energy flows through the resulting pathway decomposition of the biomolecular reaction network. Finally we apply the energy-based pathway analysis to investigate the partition of energy flows in a biomolecular cycle model of a trans-membrane transporter protein, taking as a specific example the Sodium-Glucose Transport protein SGLT1.}

\section{Energy-based Reaction Analysis}
\label{sec:energy-based-react}
\begin{figure}[htbp]
  \centering
  \SubFig{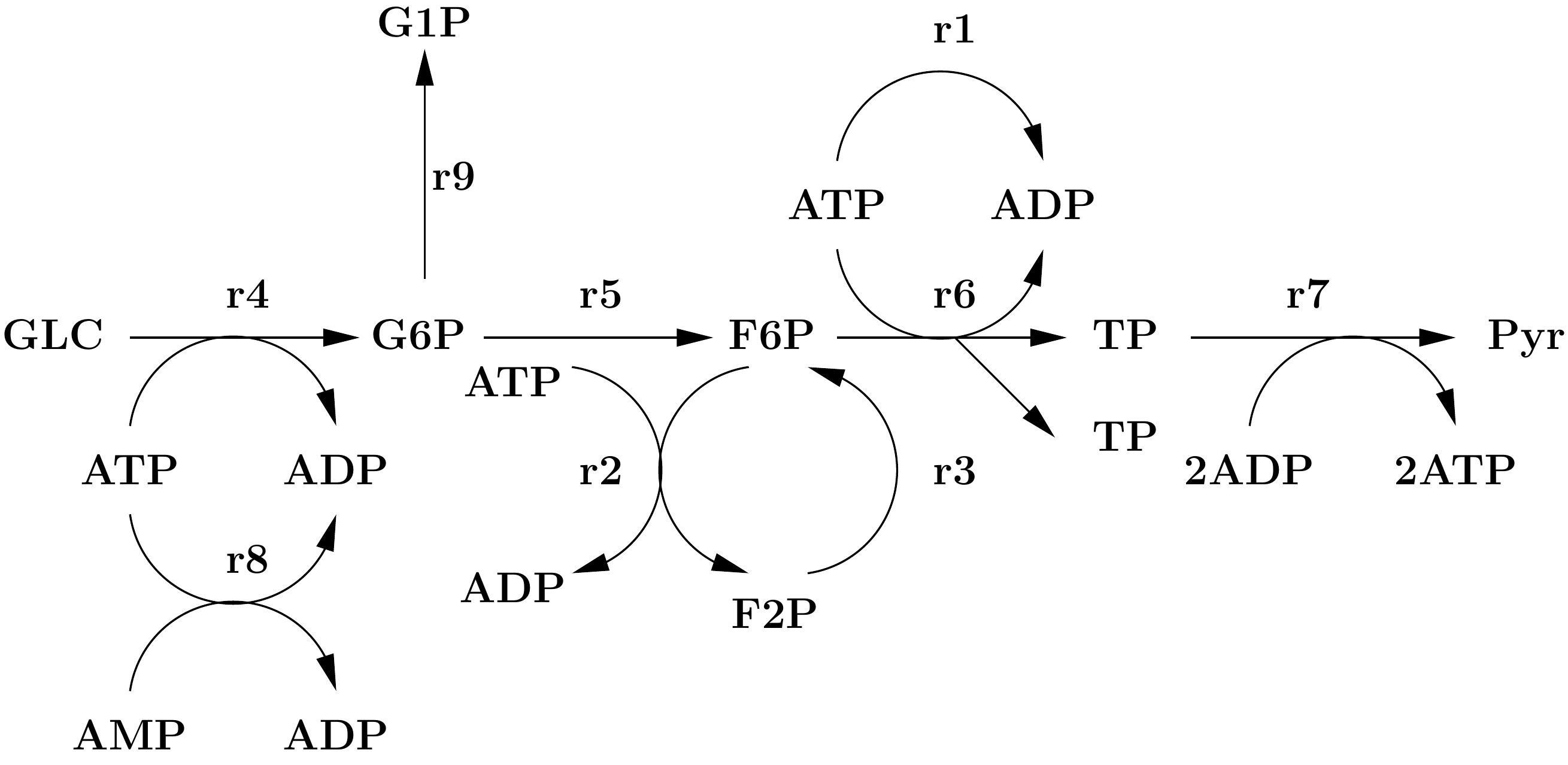}{Biomolecular reaction diagram of the glycolysis pathway}{0.6}  
  \SubFig{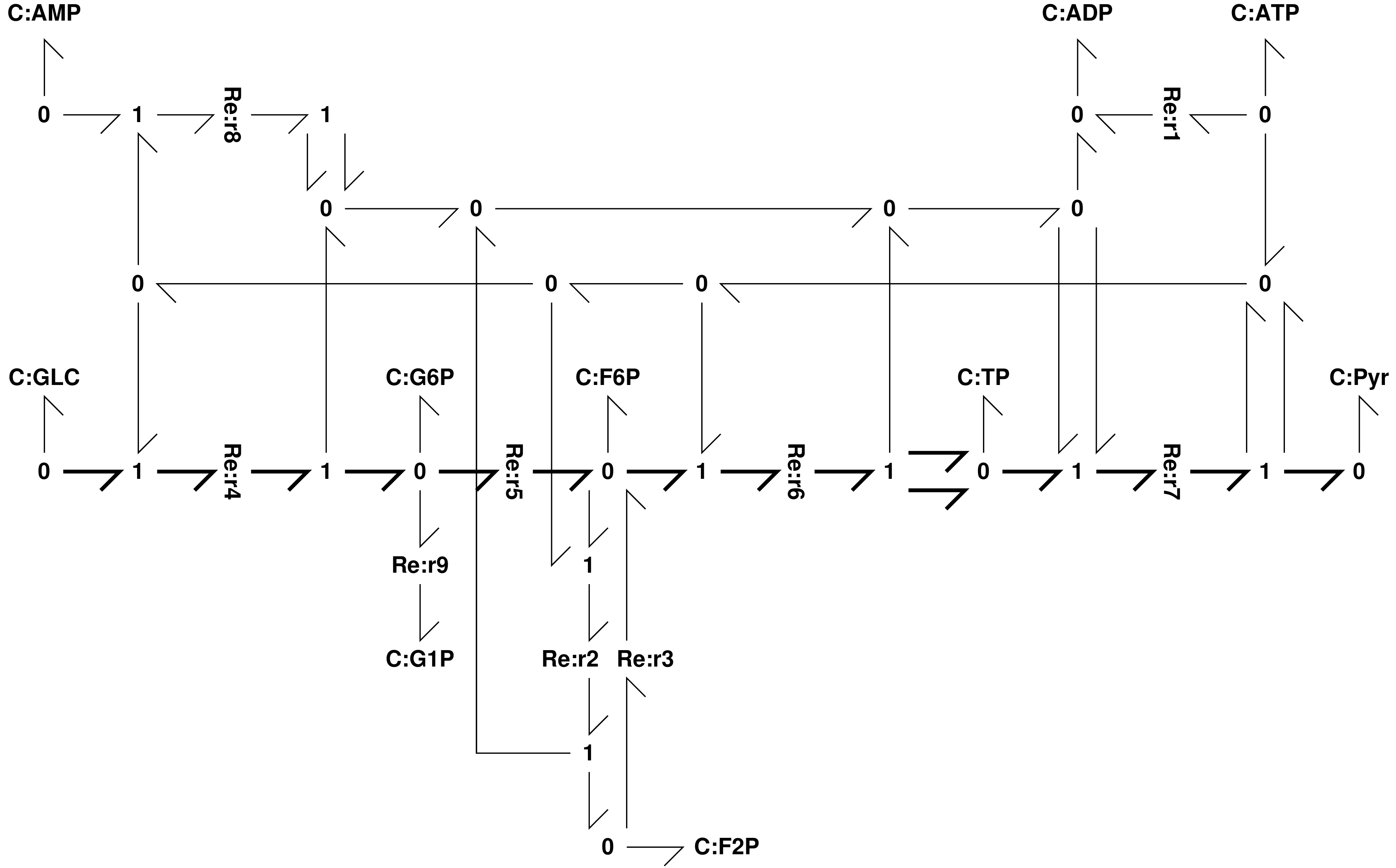}{Bond Graph representation of the biomolecular pathway shown above}{0.9}
  \caption{Glycolysis example from \citet{HeiSch96}.}
\label{fig:HeiSch96}
\end{figure}
%
% \TBD{\added{I think we need a bit more introduction to energy-based approach, motivation for it, and bond graphs.}}
% Like engineering systems, living systems are subject to the laws of
% physics in general and the laws of thermodynamics in particular.
% This fact gives the opportunity of applying energy-based approaches to
% the modelling, analysis and understanding of living systems.
%
As reviewed in the Introduction, and as discussed by
\citet{GawCra14,GawCra16} and Appendix \ref{sec:short-intr-bond},
chemical equations can be written in the form of bond graphs to enable
energy-based analysis. As an introductory and illustrative example,
the glycolysis example from the seminal book of \citet[Figure
3.4]{HeiSch96} has the following chemical equations:
\begin{xalignat*}{3}
\Sp{ATP} &\reacu{r1} \Sp{ADP} &
\Sp{ATP}+\Sp{F6P} &\reacu{r2} \Sp{ADP}+\Sp{F2P} &
\Sp{F2P} &\reacu{r3} \Sp{F6P} \\
\Sp{GLC}+\Sp{ATP} &\reacu{r4} \Sp{ADP}+\Sp{G6P} &
\Sp{G6P} &\reacu{r5} \Sp{F6P} &
\Sp{ATP}+\Sp{F6P} &\reacu{r6} \Sp{ADP}+\Sp{2TP} \\
\Sp{2ADP}+\Sp{TP} &\reacu{r7} \Sp{Pyr}+2\Sp{ATP} &
\Sp{ATP} +\Sp{AMP} &\reacu{r8} 2\Sp{ADP} &
\Sp{G6P} &\reacu{r9} \Sp{G1P}
\end{xalignat*}
These equations can be represented by the biomolecular reaction diagram of Figure
\ref{subfig:HeiSch96} or the equivalent bond graph of Figure
\ref{subfig:HeiSch96_abg}. 
Briefly, each reaction is represented by an \Re component and each
species by a \C component. The (Gibbs) energy flows are represented by
bonds $\rightharpoondown$ which carry both chemical potential and
molar flow. Bonds are connected by \zero junctions which imply a
common chemical potential on each bond and \one junctions which imply
a common molar flow on each bond. 
\C components store,  \Re components dissipate and bonds and junctions
transmit Gibbs energy.
Further details are given in Appendix \ref{sec:short-intr-bond} and references
\citep{GawCra14,GawCurCra15,GawCra16,Gaw17}.

Closed biomolecular systems are described in stoichiometric form as:
\begin{equation}
  \label{eq:dX}
\dX = \NN \VV
\end{equation}
where $X$ is the $n_X \times 1$ system state, $\VV$ the $n_V \times 1$
vector of reaction flows and and $N$ is the $n_X \times n_V$
stoichiometric matrix which can be derived from the bond graph
representation~\citep{GawCra14,Gaw17}. In the case of the example of
Figure \ref{fig:HeiSch96}:
\begin{xalignat}{3}\label{eq:HeiSch96}
X &= 
 \begin{pmatrix}
x_{GLC}\\
x_{Pyr}\\
x_{ATP}\\
x_{ADP}\\
x_{G1P}\\
x_{G6P}\\
x_{F6P}\\
x_{TP}\\
x_{F2P}\\
x_{AMP}\\
\end{pmatrix},\;&
  N &=
  \begin{pmatrix}
    0&0&0&-1&0&0&0&0&0\\
    0&0&0&0&0&0&1&0&0\\
    -1&-1&0&-1&0&-1&2&-1&0\\
    1&1&0&1&0&1&-2&2&0\\
    0&0&0&0&0&0&0&0&1\\
    0&0&0&1&-1&0&0&0&-1\\
    0&-1&1&0&1&-1&0&0&0\\
    0&0&0&0&0&2&-1&0&0\\
    0&1&-1&0&0&0&0&0&0\\
    0&0&0&0&0&0&0&-1&0
  \end{pmatrix}&
V &= 
 \begin{pmatrix}
v_{r1}\\
v_{r2}\\
v_{r3}\\
v_{r4}\\
v_{r5}\\
v_{r6}\\
v_{r7}\\
v_{r8}\\
v_{r9}\\
\end{pmatrix}
\end{xalignat}

In the case of mass-action kinetics~\citep{KliLieWie11}, the reaction flows are generated
by the formula:
\begin{equation}\label{eq:V}
  \VV = \kkappa \lb \exp\frac{\Af}{RT} - \exp\frac{\Ar}{RT}\rb
\end{equation}
where $\Af$ and $\Ar$ are the forward and reverse%
\footnote{The terms ``forward'' and ``reverse'' often correspond to
  ``substrate'' and ``product'' respectively or ``input'' and
  ``output'' respectively; they are used here for consistency with
  previous work, to avoid ambiguity and to recognise that reactions
  may be reversible.}
reaction affinities. As discussed by \citet{GawCra14,GawCra16}, these
affinities are given by:
\begin{equation}\label{eq:Afr}
  \Af = {\Nf}^T \mmu,\; \Ar = {\Nr}^T \mmu
\end{equation}
where $\text{ }^T$ indicates matrix transpose and $\mu$ is the (vector of) chemical potentials where the $i$th
element is a logarithmic function of the $i$th
element $x_i$ of $X$:
\begin{equation}\label{eq:mu}
  \mu_i = RT\ln K_i x_i
\end{equation}
where $K_i$ is a species-specific positive constant.
$\Nf$ and $\Nr$ are the forward and reverse stoichiometric matrices,
and conservation of energy requires that:
\begin{equation}\label{eq:N}
  -\Nf + \Nr = \NN
\end{equation}
%%\deleted{The conventions of chemical formulae require that} 
By definition, all
stoichiometric parameters, that is the elements $\Nf_{ij}$ of $\Nf$ and
the elements $\Nr_{ij}$ of $\Nr$ have the following properties:
\begin{xalignat}{4}
&\Nf_{ij} \text{ is integer } &&\Nf_{ij}\ge0 &
&\Nr_{ij} \text{ is integer } &&\Nr_{ij}\ge0\label{eq:Nfij}
\end{xalignat}
% \TBD{ We could express this as $\Nf_{ij} \in \mathbb{Z}_{\ge 0}$, but
%   would anyone understand it? \added{I think probably not!}}

% In this case, $\Nf$ and $\Nr$ may be obtained from the negative and
% positive elements respectively of $\NN$

As discussed by \citet{GawCra16}, open biomolecular systems can be
described and analysed using the notion of \emph{chemostats}
\citep{PolEsp14,GawCra16}. Chemostats have two biomolecular
interpretations:
\begin{enumerate}
\item one or more species are fixed to give a constant concentration (for example under a specific experimental protocol); this implies that an appropriate external
  flow is applied to balance the internal flow of the species.
\item as a \C component with a fixed state imposed on a model in order to analyse its properties~\citep{GawCurCra15}.
% \item \deleted{an ideal feedback controller is applied to species to be fixed
%   with setpoint as the fixed concentration and control signal an
%   external flow.}
\end{enumerate}
Additionally, in the context of a control systems analysis, the
chemostat can be used as an ideal feedback controller, applied to
species to be fixed with setpoint as the fixed concentration and
control signal an external flow.

When chemostats are present, the reaction flows are determined by the
dynamic part of the stoichiometric matrix. In this case the
stoichiometric matrix $N$ can be decomposed as the sum of two
matrices~\citep{GawCra16}: the \emph{chemostatic} stoichiometric matrix $N^{cs}$ and
the \emph{chemodynamic} stoichiometric matrix $N^{cd}$ where $N =
N^{cs} + N^{cd}$ and:
\begin{xalignat}{4}
  N^{cs} &= I^{cs}N &
  N^{cd} &= I^{cd}N &
  I^{cs}_{ii} &= 
  \begin{cases}
    1 & \text{if $i \in \II^{cs}$}\\
    0 & \text{if $i \not\in \II^{cs}$}
  \end{cases}&
  I^{cd}_{ii} &= 
  \begin{cases}
    0 & \text{if $i \in \II^{cs}$}\\
    1 & \text{if $i \not\in \II^{cs}$}
  \end{cases} \label{eq:N^cd}
\end{xalignat}
Note that $N^{cd}$ is the same as $N$ except that the \emph{rows}
corresponding to the chemostat variables are set to zero. In this
case $N^{cd}$ is given by $N$ of Equation \eqref{eq:HeiSch96} with rows
$1$--$4$ set to zero. 

With these definitions, an open system can be expressed as
\begin{equation}
  \label{eq:open}
\dX = N^{cd} \VV
\end{equation}
The stoichiometric properties of $N^{cd}$, rather than $N$, determine
system properties when chemostats are present.
Using equation \eqref{eq:N^cd}, $\dX$ in equation \eqref{eq:open} can also be
written as:
\begin{xalignat}{2}
  \dX &= N \VV -  N^{cs}\VV = N \VV + \VV^s &
\text{where } \VV^s &=  -N^{cs}\VV
\end{xalignat}
$V^s$ can be interpreted as the external flows required to hold the
chemostat states constant. The reaction flows are given by  the same
formulae \eqref{eq:V}~\&~\eqref{eq:Afr} as closed systems.

\section{Stoichiometric Pathway Analysis}
\label{sec:stoich-analys-pathw}
As discussed in the textbooks
\citep{Cla88,HeiSch96,Pal06,Pal11,KliLieWie11} and Appendix
\ref{sec:short-intr-syst}, the (non-unique) $n_V \times n_P$
null-space matrix $\Kp$ of the open system stoichiometric matrix
$\Ncd$ has the property that
\begin{xalignat}{2}\label{eq:NK}
  \Ncd \Kp &= 0 &
\text{where } n_P &= n_V-r
\end{xalignat}
and $r$ is the rank of $\NN$.
Furthermore, if the reaction flows $\VV$ are constrained in terms
of the $n_P$ pathway flows $\Vp$ as
\begin{equation}\label{eq:KV}
  \VV = \Kp \Vp
\end{equation}
then substituting Equation \eqref{eq:KV} into Equation \eqref{eq:dX} and
using Equation \eqref{eq:NK} implies that $\dX=0$. This is
significant because the biomolecular system of equation \eqref{eq:dX}
may be in a steady state for any choice of $\Vp$.

As mentioned above, $\Kp$ is not unique: there are many possible
approaches to choosing $\Kp$ such that Equation \eqref{eq:NK}
holds. As discussed by, for example, \citet{PfeSanNun99}, $\Kp$ can be
computed in such a way as to give useful features such as integer
entities and maximal number of zero elements; moreover, if all
reactions are irreversible, the columns of $\Kp$ must correspond to a
convex space%
\footnote{This is related to the classical Cone
  Lemma~\citep[Chap. 10]{AigZie14}.}
. 
As discussed in \S\ref{sec:pathway-analysis}, the analysis of this
paper requires that all elements $\Kp_{ij}$ of $\Kp$ must satisfy the
same conditions as those on $\Nf_{ij}$ and $\Nr_{ij}$
\eqref{eq:Nfij}, namely:
\begin{xalignat}{2}
&\Kp_{ij} \text{ is integer} &\Kp_{ij}\ge0\label{eq:Kpij}
\end{xalignat}
For this reason, $\Kp$ is referred to as the \emph{positive-pathway
  matrix} (PPM) in the sequel.
This does \emph{not} imply that reactions are assumed to be
irreversible; however, one approach to generating such a $\Kp$ is
using software such as \texttt{metatool} \citep{PfeSanNun99} as if all
reactions were irreversible.

% \TBD{Existence and uniqueness results. Eg. \citet{SchHilWooFel02},
%   ``Cone Lemma'' \citet[Chap. 10]{AigZie14} \& Ivo's email. \added{probably not necessary?}}

In the case of the glycolysis system of Figure \ref{fig:HeiSch96},
\citet{HeiSch96} choose the chemostats to be: $\Sp{GLC}$, $\Sp{Pyr}$,
$\Sp{ATP}$ and $\Sp{ADP}$. Using the algorithm of
\citet{PfeSanNun99}%
\footnote{The Octave~\citep{EatBatHau15}  version of metatool from
  \url{http://pinguin.biologie.uni-jena.de/bioinformatik/networks/metatool/metatool5.0/metatool5.0.html}
was used.},
 $\Kp$ for the glycolysis system of Figure
\ref{fig:HeiSch96} was computed as
%%\citep[(3.4.9)]{HeiSch96} as: 
%%\TBD{\added{indicate how $\Kp$ is calculated? (using 3.1 and 3.4)}}
%
\begin{equation}\label{eq:Kp_HeiSch96}
\Kp =
\begin{pmatrix}
1&0&0\\
0&1&0\\
0&1&0\\
0&0&1\\
0&0&1\\
0&0&1\\
0&0&2\\
0&0&0\\
0&0&0
\end{pmatrix}
\end{equation}
The pathways corresponding to the three columns of $\Kp$, (referring to Figure \ref{subfig:HeiSch96}), are: 
$\begin{Bmatrix} r1 \end{Bmatrix}$,
$\begin{Bmatrix} r2&r3 \end{Bmatrix}$ and
$\begin{Bmatrix} r4& r5& r6& 2r7 \end{Bmatrix}$.
The latter pathway is marked on the bond graph  of
\ref{subfig:HeiSch96_abg} using bold bonds. 
As pointed out by \citet{HeiSch96}, these three pathways are
independent insofar as they have no reactions in common; as will be
seen in the rest of this section and in the examples of
\S\ref{sec:example:-free-energy} and \S\ref{sec:exampl-sodi-gluc},
this is not always the case.

The number of pathways, and their independence, not only depend on the
network structure, but also on the choice of chemostats. To illustate
this point, consider the same glycolysis system of Figure \ref{fig:HeiSch96},
 but choose an additional chemostat to be $\Sp{G1P}$.  In this case,
 $\Kp$ is given
by:
\begin{equation}\label{eq:Kp_HeiSch96_G1p}
\Kp =
\begin{pmatrix}
1&0&0&0\\
0&1&0&0\\
0&1&0&0\\
0&0&1&1\\
0&0&1&0\\
0&0&1&0\\
0&0&2&0\\
0&0&0&0\\
0&0&0&1
\end{pmatrix}
\end{equation}
The first three columns of Equation \eqref{eq:Kp_HeiSch96_G1p} are
identical to the three columns of Equation \eqref{eq:Kp_HeiSch96}. The
fourth column corresponds to the additional pathway 
$\begin{Bmatrix} r4 & r9 \end{Bmatrix}$,
which shares reaction $r4$ with the pathway 
$\begin{Bmatrix} r4& r5& r6& 2r7 \end{Bmatrix}$
coresponding to the third column of Equations
\eqref{eq:Kp_HeiSch96_G1p} and \eqref{eq:Kp_HeiSch96}.
%
%% \TBD{\added{Perhaps this next paragraph as a footnote? Not sure it
%% will be meaningful to reader (unless already familiar with
%% causality etc)}}
%
% \footnote{An alternative approach to generating a meaningful $\Kp$ is via
% causality analysis of the bond graph  describing the system
% \citep{GawCra14,Gaw17}.
% %
% Using the Bond Graph based causality analysis of \citet{Gaw17}, the
% vector of independent flows $\Vp$ is chosen by the modeller and can be
% written in the form:
% \begin{equation}
%   \Vp = T_I \VV
% \end{equation}
% The corresponding $\Kp$ is then uniquely defined.}.

% This paper is not concerned with how a meaningful $\Kp$ is generated,
% but rather how an energy-based analysis may be combined with a pathway
% analysis based on $\Kp$ to yield additional insights into biomolecular
% systems.

\section{Energy-based Pathway Analysis}
\label{sec:pathway-analysis}
\begin{figure}[htbp]
  \centering
  \SubFig{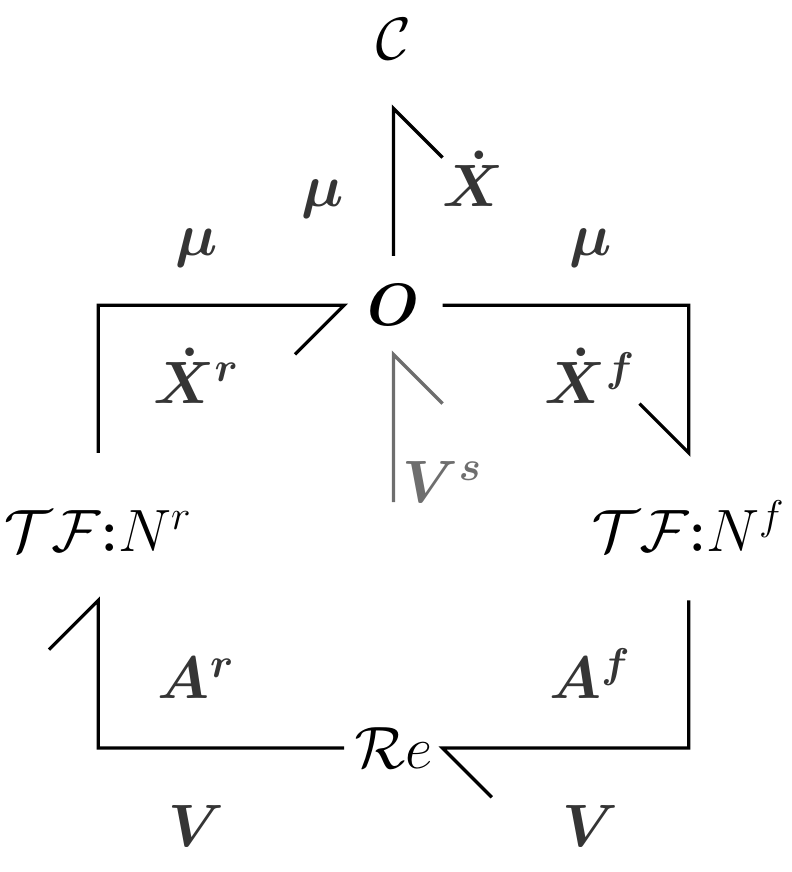}{Reaction-based}{0.3}
  \SubFig{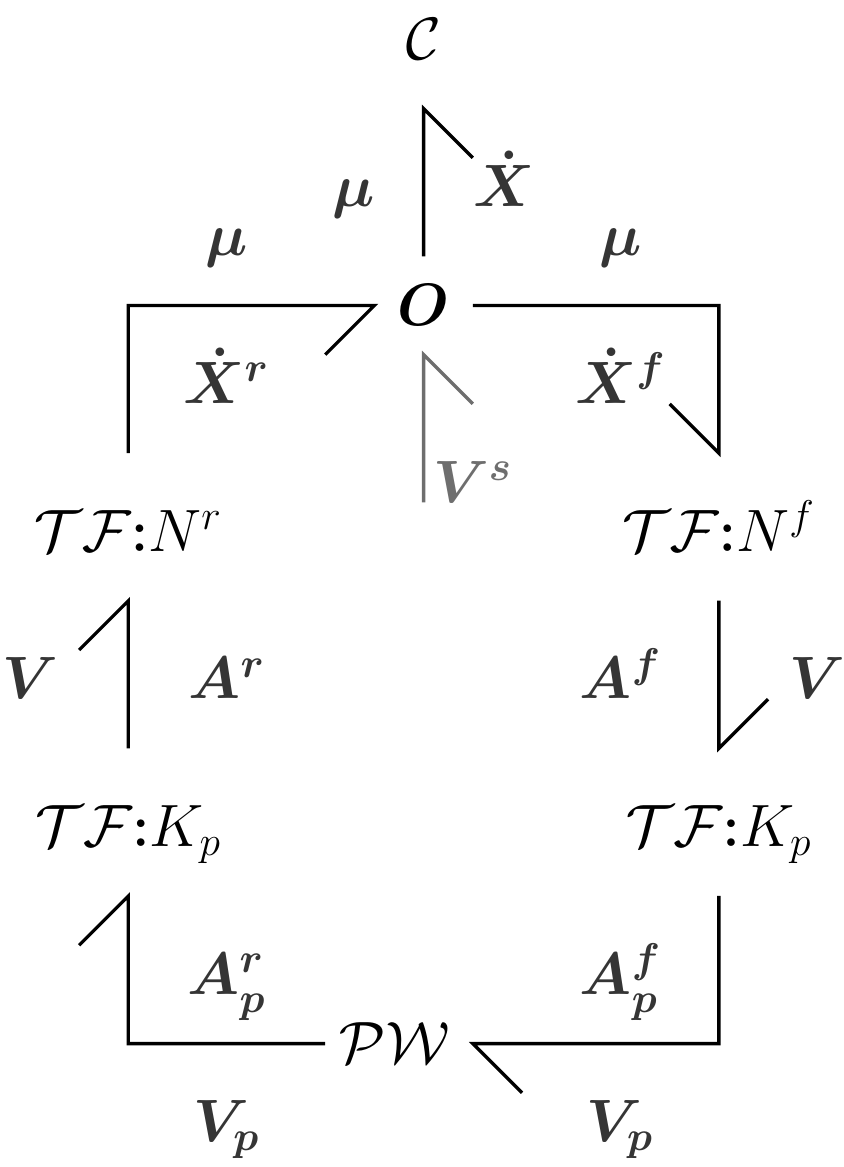}{Pathway-based}{0.3}
  \SubFig{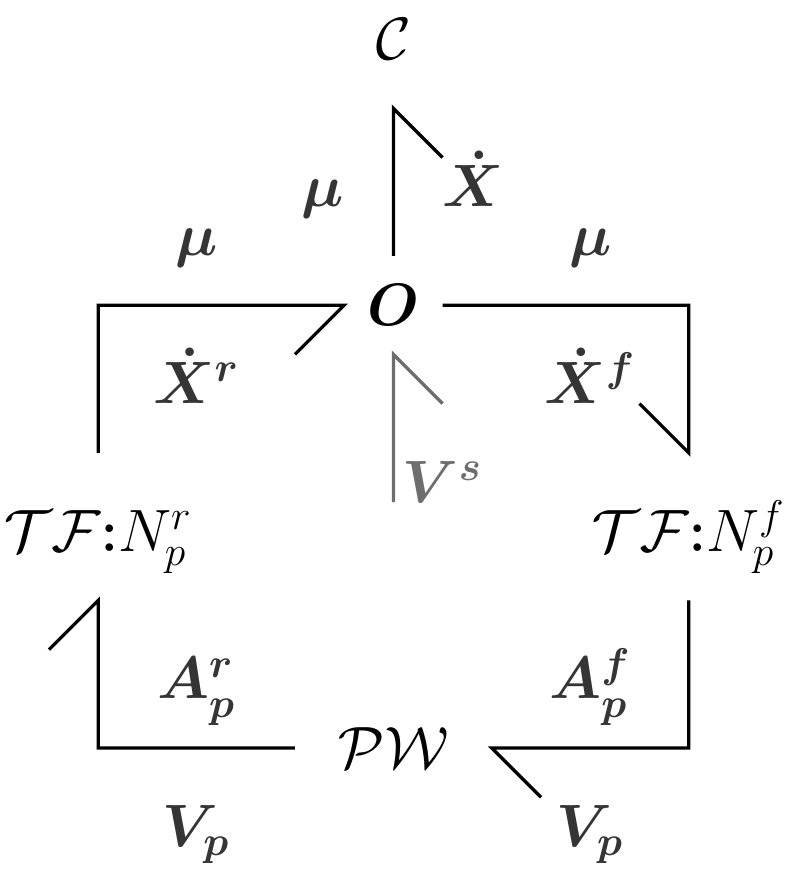}{Pathway-based}{0.3}
  \caption{Energy-based analysis. (a) Reaction-based system
    \citep{GawCra16}. The bond symbols $\rightharpoondown$ correspond
    to \emph{vectors} of bonds; $\mathcal{C}$, $\mathcal{R}e$ and
    $\mathcal{O}$ correspond to arrays of \C, \Re and \zero components
    ; the two $\mathcal{TF}$ components represent the intervening
    junction structure comprising bonds, \zero and \one junctions and
    \TF components. $N^f$ and $N^r$ are the forward and reverse
    stoichiometric matrices. $V^s$ represents the chemostatic
    flows. (b) The positive-pathway matrix $\Kp$ maps the $n_P$
    positive-pathway flows $\Vp$ onto the $n_V$ reaction flows $\VV$
    and also maps the $n_V$ reaction affinities onto the $n_P$
    positive-pathway affinities. $\PW$ conceptually
    represents the non-linear function generating the steady-state
    positive-pathway flows in terms of the positive-pathway
    affinities; unlike $\mathcal{R}e$, it does \emph{not} have a diagonal
    structure. (c) $\mathcal{TF}$ components have been merged using
    $\Npf = \Nf\Kp$ and $\Npr = \Nr\Kp$.}
\label{fig:EBA}
\end{figure}
This section combines the energy-based reaction analysis of
\S\ref{sec:energy-based-react} with the stoichiometric pathway
analysis of \S\ref{sec:stoich-analys-pathw} to give an energy-based
pathway analysis of biomolecular systems.

As discussed by \citet{GawCra14,GawCra16}, Equations \eqref{eq:dX},
\eqref{eq:Afr} and \eqref{eq:N} can be summarised by the diagram of
Figure \ref{subfig:Open_bg}. The key point here is that the energy
flow from the reaction components represented by the vectors of
covariables $\Ar$ and $\VV$ is transformed without energy loss by
$\mathcal{TF}:N^r$ into the energy flow to the $\mathcal{C}$
components represented by the vectors of covariables $\mmu$ and
$\dX^r$ and the energy flow from the $\mathcal{C}$ components
represented by the vectors of covariables $\mmu$ and $\dX^f$ is
transformed without energy loss by $\mathcal{TF}:N^f$ into the energy
flow to reaction components represented by the vectors of covariables
$\Af$ and $\VV$.  Dissipation of energy occurs at the reaction
components $\mathcal{R}e$.  The challenge for an energy-based pathway
analysis is to determine the energy flow associated with each pathway.

The key result of the stoichiometric pathway analysis summarised in
\S\ref{sec:stoich-analys-pathw} is Equation \eqref{eq:KV} relating the
pathway flow vector $\Vp$ to the reaction flow vector $\VV$ by
$\VV = \Kp \Vp$. To bring this result into the energy domain, we
define the forward and reverse pathway affinities $\Apf$ and $\Apr$,
and the pathway affinity $\Ap$, as:
\begin{xalignat}{3}\label{eq:A_P}
  \Apf &= {\Kp}^T \Af &
  \Apr &= {\Kp}^T \Ar &
  \Ap &= \Apf-\Apr = {\Kp}^T \lb \Af-\Ar\rb = {\Kp}^T\AA 
\end{xalignat}
With these definitions we can define powers (energy flows) associated
with pathways:
\begin{align}
  \Ppf &= \Apf^T\Vp = \lb {\Kp}^T \Af \rb^T\Vp = \Af^T\Kp\Vp =
         \Af^T\VV = \Pf\\
\text{similarly }   \Ppr &= \Apr^T\Vp = \lb {\Kp}^T \Ar \rb^T\Vp = \Ar^T\Kp\Vp =
         \Ar^T\VV = \Pr
\end{align}
Thus the net forward pathway energy flow $\Ppf$ equals the net forward
reaction energy flow $\Pf$ and the net reverse pathway energy flow
$\Ppr$ equals the net reverse reaction energy flow $\Pr$.  Hence $\Kp$
can be considered as an energy transmitting transformer for both
forward and backward pathway energies and is represented in Figure
\ref{subfig:Open_SS_bg} by the symbol $\mathcal{TF}:\Kp$.
%
%%\deleted{Note that as }
 
Figure \ref{subfig:Open_SS_bg_1} can be obtained from Figure
\ref{subfig:Open_SS_bg} by combining Equations \eqref{eq:dX} and
\eqref{eq:N} with Equation \eqref{eq:KV}, and Equations \eqref{eq:A_P}
with Equations \eqref{eq:Afr} to give:
\begin{xalignat}{3}
  \dX &= \Np \Vp &
  \Apf &= \Npf^T \mmu&
  \Apr &= \Npr^T \mmu\label{eq:Apfr}\\
\text{where }
\Np &= \NN \Kp &
\Npf &= \Nf \Kp &
\Npr &= \Nr \Kp \label{eq:NP} 
\end{xalignat}
%%\TBD{\added{These define $\Npf$ and $\Npr$ (I think?)}}
Equation \eqref{eq:NP} defines the forward $\Npf$ and reverse $\Npr$
pathway stoichiometric matrices;
using conditions \eqref{eq:Nfij} and \eqref{eq:Kpij},
it follows that $\Npf$ and $\Npr$ have positive integer elements:
\begin{xalignat}{4}
&\Npf_{ij} \text{ is integer} &&\Npf_{ij}\ge0&
&\Npr_{ij} \text{ is integer} &&\Npr_{ij}\ge0\label{eq:Npfij}
\end{xalignat}

A property of $\Np$ follows from combining Equations \eqref{eq:NP},
\eqref{eq:N^cd} and \eqref{eq:NK}; in particular, $\Np$ may be rewritten as
\begin{equation}
  \Np = \lb N^{cs} + N^{cd} \rb \Kp = N^{cd} \Kp = I^{cs} \NN \Kp
\end{equation}
Hence $\Np$ has the property that the only non-zero rows correspond to
the chemostats.

In the case of the example of Figure \ref{fig:HeiSch96}:
\begin{xalignat}{3}\label{eq:N_HeiSch96}
  \Np &=
  \begin{pmatrix}
0&0&-1\\
0&0&2\\
-1&-1&2\\
1&1&-2\\
0&0&0\\
0&0&0\\
0&0&0\\
0&0&0\\
0&0&0\\
0&0&0
  \end{pmatrix}&
  \Npf &=
  \begin{pmatrix}
0&0&1\\
0&0&0\\
1&1&2\\
0&0&4\\
0&0&0\\
0&0&1\\
0&1&1\\
0&0&2\\
0&1&0\\
0&0&0
  \end{pmatrix}&
  \Npr &=
  \begin{pmatrix}
0&0&0\\
0&0&2\\
0&0&4\\
1&1&2\\
0&0&0\\
0&0&1\\
0&1&1\\
0&0&2\\
0&1&0\\
0&0&0
  \end{pmatrix}
\end{xalignat}

Using Equation \eqref{eq:Apfr} and $\Np$ from Equation
\eqref{eq:N_HeiSch96}, the affinity associated with each of the three
pathways is, as expected:
\begin{xalignat}{2}
{\Ap}_1 &= {\Ap}_2 = \mmu_{ATP} - \mmu_{ADP}&
{\Ap}_3 &= \lb \mmu_{GLC} + 2\mmu_{ADP}\rb - 
\lb 2\mmu_{Pyr} + 2\mmu_{ATP}\rb
\end{xalignat}

Because conditions \eqref{eq:Npfij} agree with
conditions \eqref{eq:Nfij}, $\Npf$
and $\Npr$ could correspond to the forward and reverse stoichiometric
matrices of the following three reactions:
\begin{align*}
  \Sp{ATP} &\reacul{p1}{PP} \Sp{ADP}\\
  \Sp{ATP} + \Sp{CS2} &\reacul{p2}{PP} \Sp{ADP}+ \Sp{CS2} \\
  \Sp{GLC}+2\Sp{ADP}+ \Sp{CS3} &\reacul{p3}{PP} 2\Sp{Pyr}+2\Sp{ATP}+\Sp{CS3}
\end{align*}
where $\Sp{CS2} = \Sp{F6P}+\Sp{F2P}$ and $\Sp{CS3} = 2\Sp{ATP} + 2\Sp{ADP}+\Sp{G6P}+\Sp{F6P}+2\Sp{TP}$.
However, some care must be taken in this interpretation. 
Firstly, the kinetics are not mass-action even if the original
equations correspond to mass-action kinetics.
Secondly, and unlike conventional reactions, these three reactions are
not independent; the consequences of this fact are discussed in
\S\ref{sec:example:-free-energy}. 
Nevertheless, such a pathway decomposition provides insight into the
energy flows in a reaction network. This is illustrated using a generic
example of \citet{Hil89} in \S~\ref{sec:example:-free-energy} and using
a specific example, the  sodium-glucose transport protein of
\citet{EskWriLoo05}, in \S~\ref{sec:exampl-sodi-gluc}.

% \TBD{Expand on this representation of pathways as reactions. Note that
%   \begin{enumerate}
%   \item the kinetics are \emph{not} mass action.
%   \item the reactions are \emph{not} independent if the pathways
%     intersect.
%   \item See \S\ref{sec:example:-free-energy} for some consequences of this.
%   \end{enumerate}}

% \TBD{\added{I think we should emphasise somewhere that the
% application to glycolysis and transporter model are illustrative
% (and informative) but this approach can be applied to arbitrarily
% 'complicated' reaction networks}}
%%% Done in conclusion.

\section{Example: Free Energy Transduction and Biomolecular Cycles}
\label{sec:example:-free-energy}
\begin{figure}[htbp]
  \centering
  \SubFig{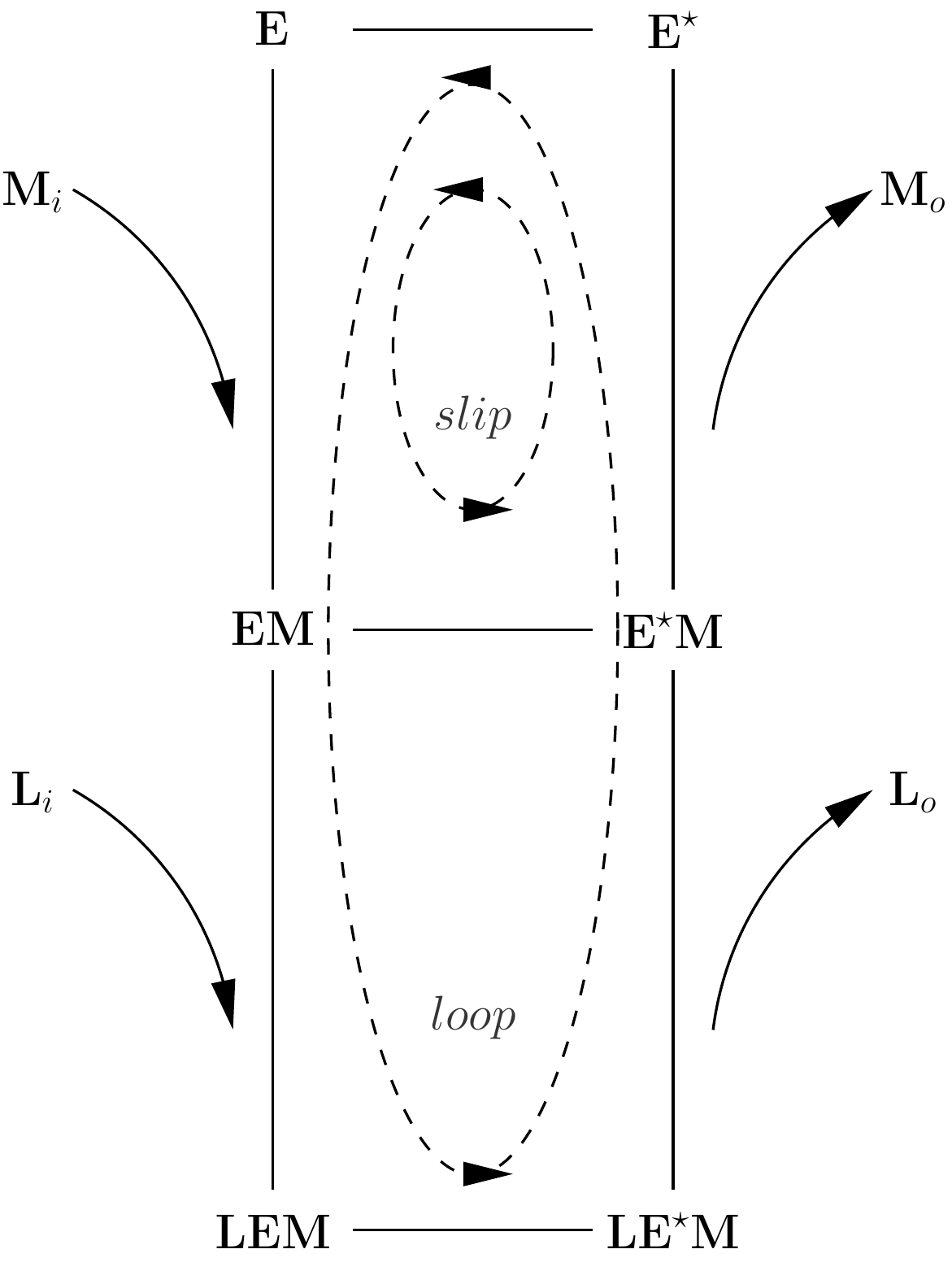}{Biomolecular cycle}{0.32}  
  \SubFig{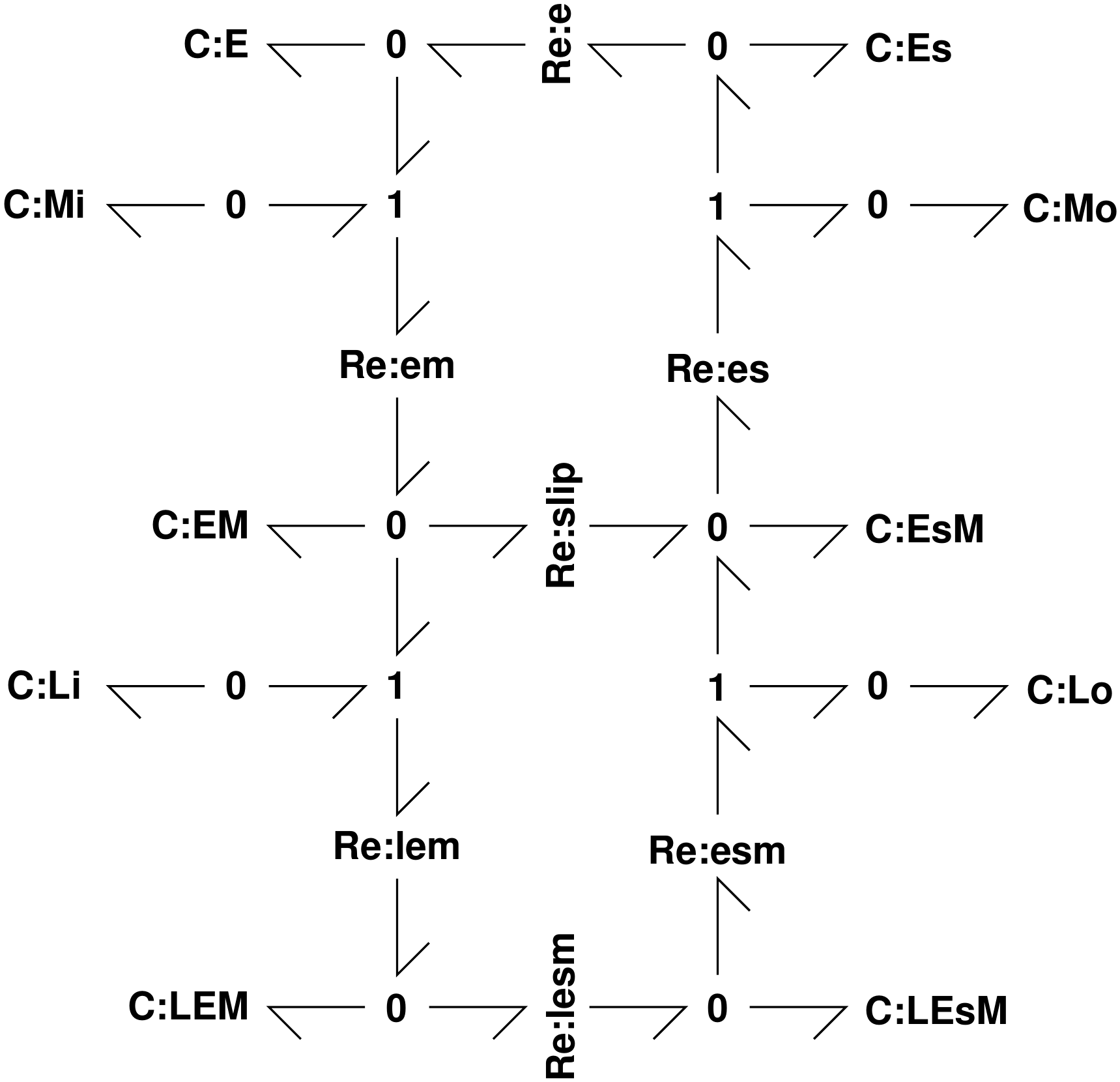}{Bond Graph representation}{0.45}  
%%  \SubFig{Hilljs_abg}{Junction structure}{0.45}
  \caption{Free energy transduction and biomolecular cycles: model. 
    (a) As discussed in the text, this is a generic model of a
    transmembrane transporter due to \citet{Hil89} and based on the 
    conformations of the
    protein $\Sp{E}$ which uses the chemical gradient of $\Sp{M}$ to
    pump $\Sp{L}$ across the membrane against an adverse gradient.
    An ideal cycle would have no ``slippage'': the link from $\Sp{EM}$ to $\Sp{E^\star M}$ would not exist. 
    As discussed in the text, two pathways: \emph{loop} and
    \emph{slip} have been marked.
    (b) The bond graph is geometrically the same as (a) but gives a
    precise description. (Es is used in place of $\Sp{E}^\star$ for
    syntactical reasons). }
\label{fig:hill}
\end{figure}
% \TBD{\added{Update these figures to use E$^*$ rather than Es? Also,
% perhaps add labels 'r1' 'r2' etc to (a)? as for the earlier example,
% and ?}}

In his classic monograph, ``Free energy transduction and biomolecular
cycle kinetics'' \citet{Hil89} discusses how the difference between
the concentration of a species $\Sp{M}_o$ outside a membrane and
the concentration of the same species $\Sp{M}_i$ inside the membrane
can be used to transport another species $\Sp{L}_i$ inside the
membrane to outside the membrane $\Sp{L}_o$ via a large protein
molecule with two conformations $E$ and $E^\star$ the former allowing
successive binding to $\Sp{M}_i$ and $\Sp{L}_i$ and the latter to
$\Sp{M}_o$ and $\Sp{L}_o$.  This biomolecular cycle is represented by
the diagram of Figure \ref{subfig:Hill_diagram} which corresponds to
that of \citet[Figure 1.2(a)]{Hil89}. There are seven reactions
 \begin{xalignat*}{4}
   \Sp{M}_i + \Sp{E} &\reacu{em} \Sp{EM} &
   \Sp{L}_i + \Sp{EM} &\reacu{lem} \Sp{LEM}&
   \Sp{LEM} &\reacu{lesm} \Sp{LE^\star M}&
   \Sp{LE^\star M} &\reacu{esm} \Sp{L}_o + \Sp{E^\star M}\\
   \Sp{E^\star M} &\reacu{es} \Sp{M}_o + \Sp{E^\star}&
   \Sp{E^\star} &\reacu{e} \Sp{E}  \label{eq:Hill-6}&
   \Sp{EM} &\reacu{slip} \Sp{E^\star M}
 \end{xalignat*}
 where the last reaction is the so-called \emph{slippage}%
\footnote{The term ``slippage'' was used in this context by Terrell L
  Hill in his seminal book~\citep{Hil89}. An ideal cycle would have no
slippage and the link from $\Sp{EM}$ to $\Sp{E^\star M}$ would not exist
in Figure \ref{subfig:Hill_diagram}.}
 term in which the enzyme changes conformation without transporting
 species $\Sp{L}$.  The original Hill diagram does not name the seven
 reactions, they are named here to provide a link to the bond graph
 which explicitly names reactions.
 % \TBD{\added{Notation E and E* different to Notation in Figure 3 (b)
 % (E and Es) - would be better to have consistent notation?}}
 
 The corresponding bond graph appears in Figure \ref{subfig:HillC_abg}
 where $\Sp{E^\star}$ is replaced by $\Sp{Es}$ for syntactical
 reasons. The bond graph clearly shows the cyclic structure of the
 chemical reactions and is
 geometrically similar to the diagram of Figure
 \ref{subfig:Hill_diagram}.
 As discussed by \citet{Hil89}, the four species $\Sp{M}_o$, $\Sp{M}_i$, $\Sp{L}_o$
 and $\Sp{L}_i$, are assumed to have constant concentration: therefore they
 are modelled by four chemostats.
\begin{figure}[htbp]
  \centering
  \SubFig{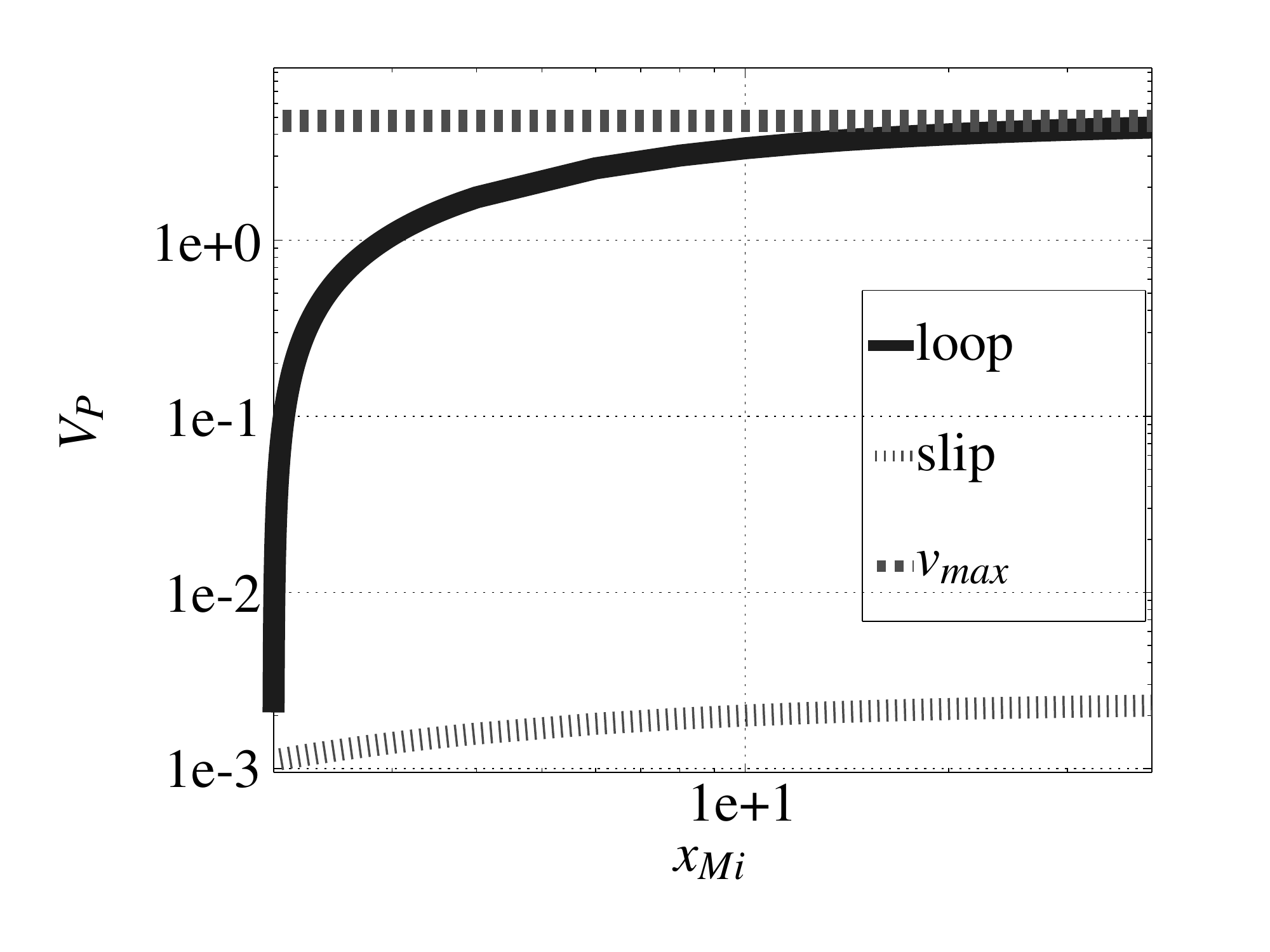}{$\kappa_{slip} = 10^{-3}$}{0.45}
  \SubFig{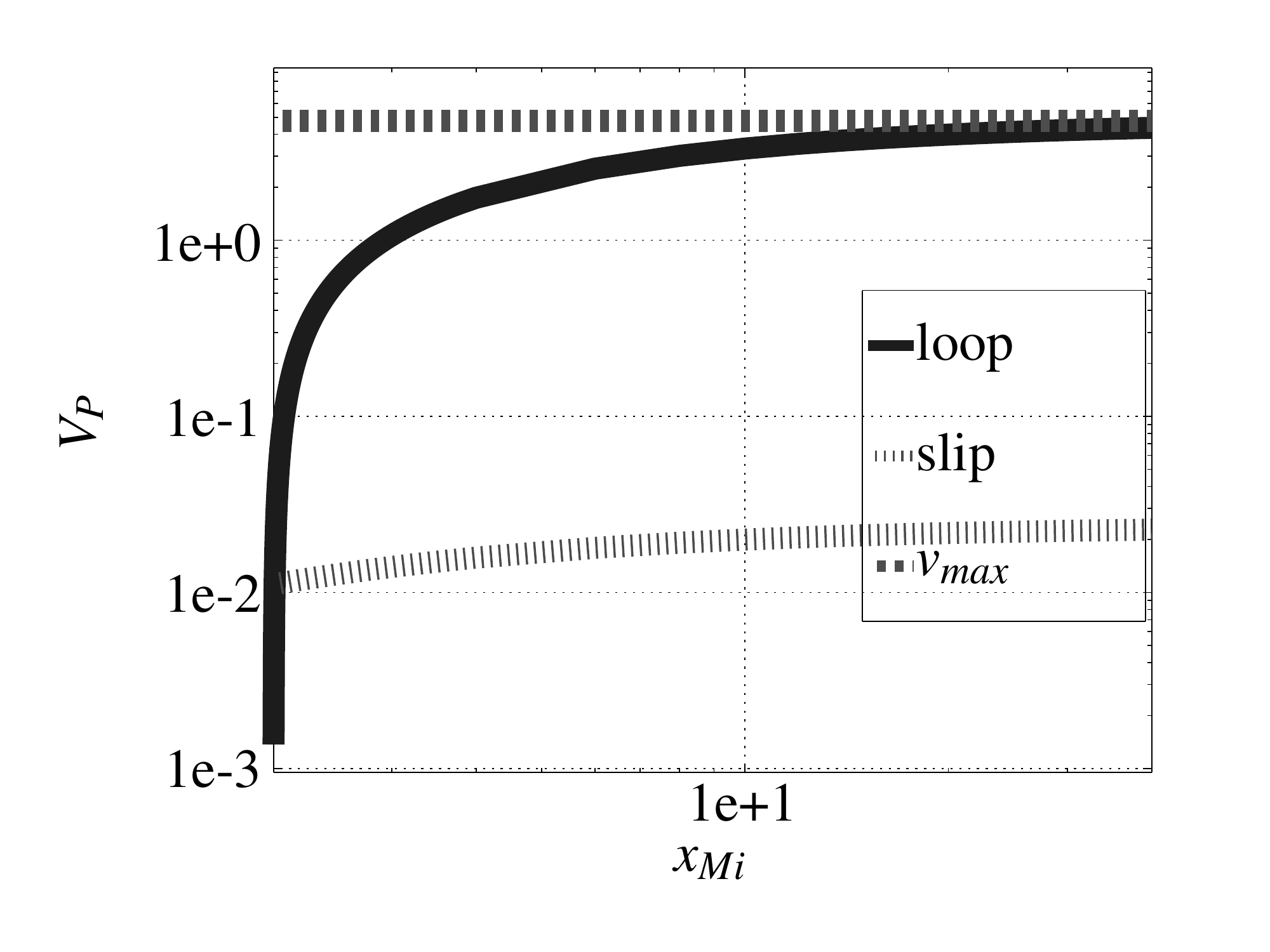}{$\kappa_{slip} = 10^{-2}$}{0.45}
  \SubFig{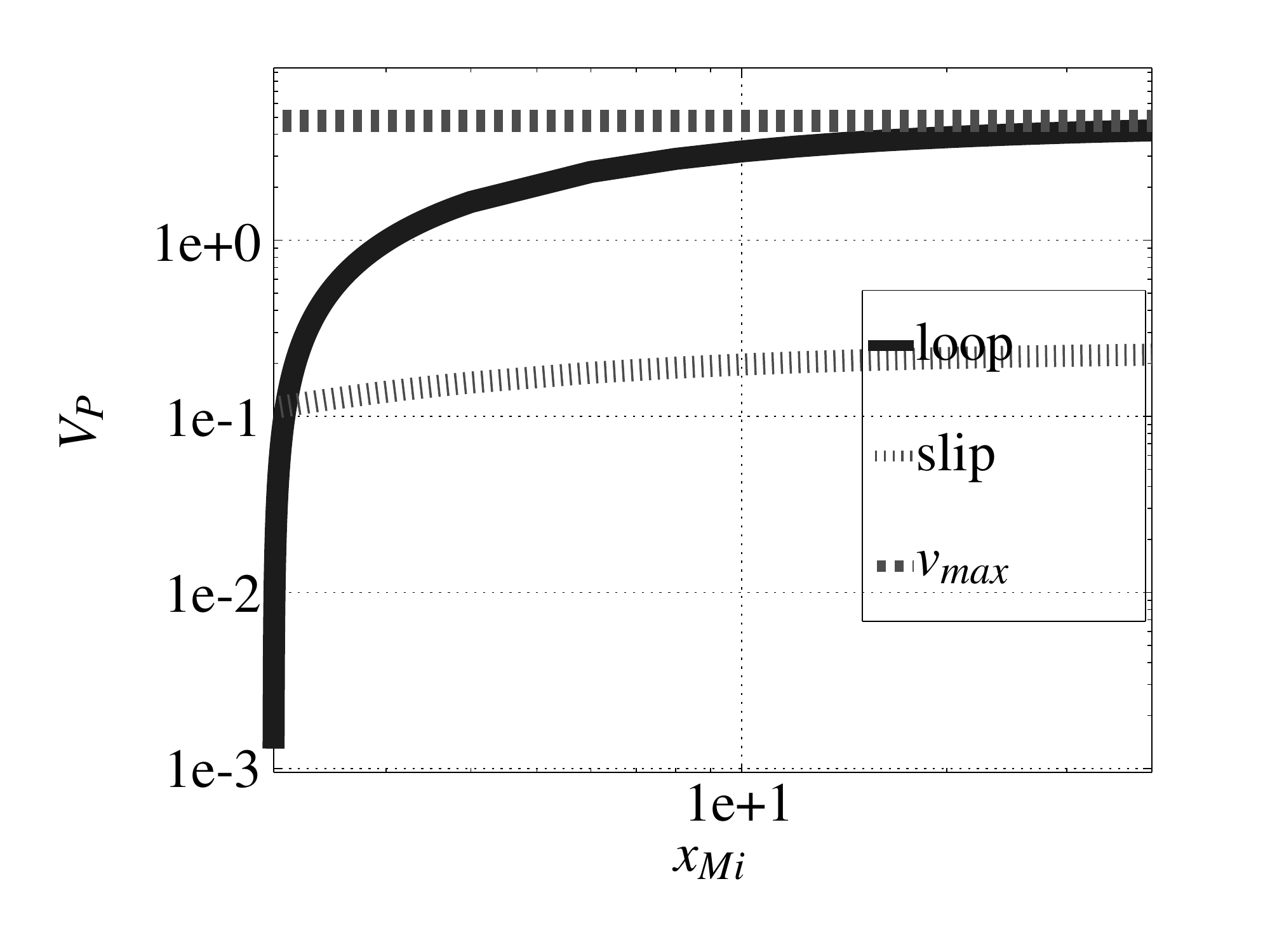}{$\kappa_{slip} = 10^{-1}$}{0.45}
  \SubFig{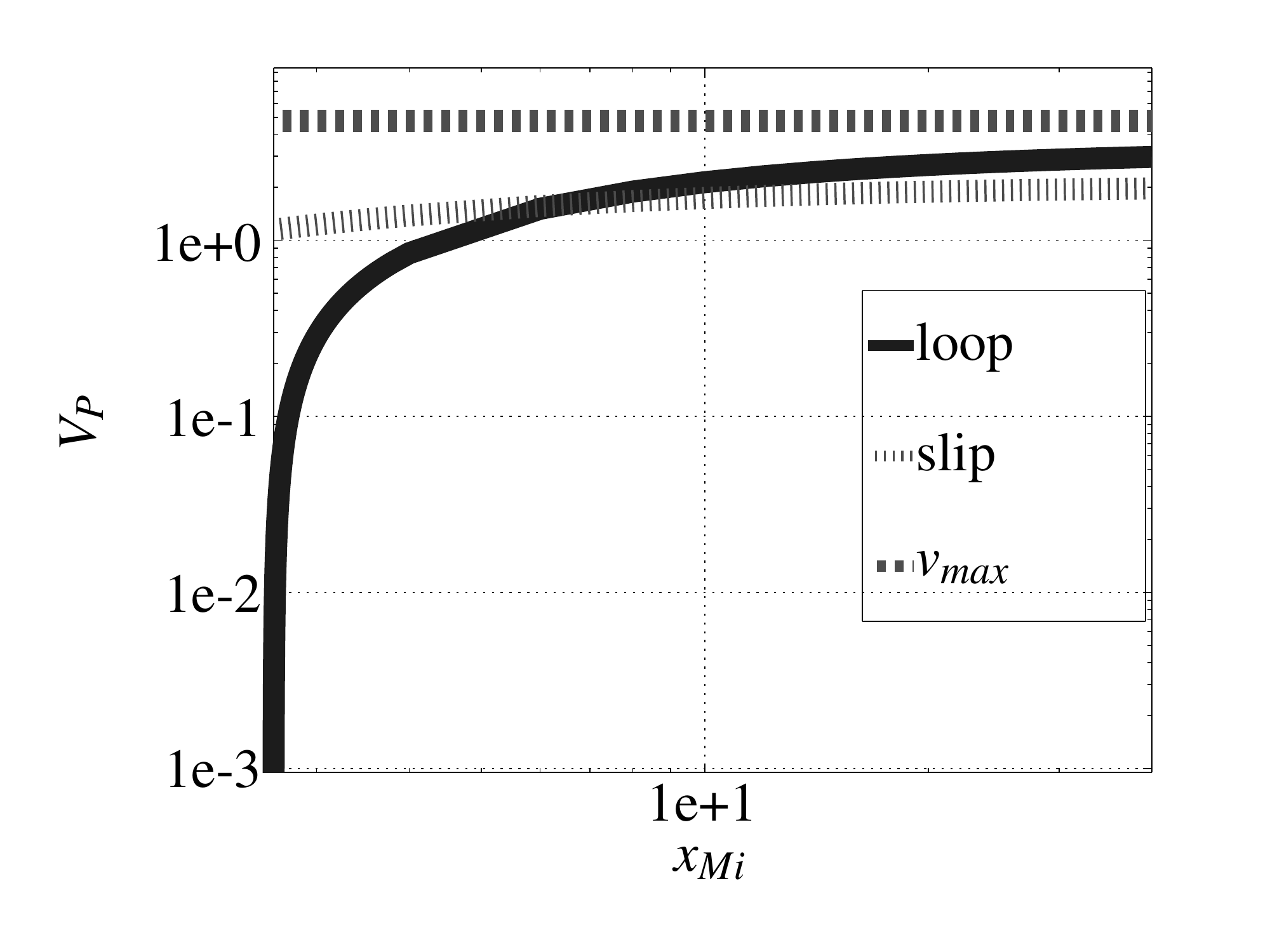}{$\kappa_{slip} = 1$}{0.45}
  \caption{Free energy transduction and biomolecular cycles: positive
    flow regime.  (a) The flow in the two pathways labelled
    \emph{loop} and \emph{slip} in Figure \ref{subfig:Hill_diagram}
    are shown together with the maximum flow $v_{max}$ for slippage
    coefficient $\kappa_{slip}=10^{-3}$. 
    (b)~--~(d) show the same information for three further values of
    the slippage coefficient $\kappa_{slip}$.}
\label{fig:hill_SS_path_pos}
\end{figure}

% \TBD{\added{Should we give the order of species (otherwise hard to
% work out which row / col refers to whcih species / reaction) and I'm
% not quite sure what order you have put the species and reactions. }}
The states $X$, stoichiometric matrix $N$ and reaction flows $V$ of
this system are:
\begin{equation}\label{eq:N_hill}
  X = 
 \begin{pmatrix}
x_{L_i}\\
x_{L_o}\\
x_{M_i}\\
x_{M_o}\\
x_{E}\\
x_{EM}\\
x_{LEM}\\
x_{E^\star}\\
x_{E^\star M}\\
x_{LE^\star M}
\end{pmatrix}, \quad
   N =
  \begin{pmatrix}
    0&-1&0&0&0&0&0\\
    0&0&0&1&0&0&0\\
    -1&0&0&0&0&0&0\\
    0&0&0&0&1&0&0\\
    -1&0&0&0&0&1&0\\
    1&-1&0&0&0&0&-1\\
    0&1&-1&0&0&0&0\\
    0&0&0&0&1&-1&0\\
    0&0&0&1&-1&0&1\\
    0&0&1&-1&0&0&0
  \end{pmatrix}, \quad
  V = 
 \begin{pmatrix}
v_{em}\\
v_{lem}\\
v_{lesm}\\
v_{esm}\\
v_{es}\\
v_{e}\\
v_{slip}
\end{pmatrix}
\end{equation}
The matrix $N^{cd}$ is the same as $N$ but with the first four rows
(corresponding to the four chemostats) deleted.  The rank of $N^{cd}$
is $r=5$ coresponding to the four chemostats and the conserved moiety
of the remaining states; thus the null space has dimension
$n_V-r = 2$.

% \TBD{\added{Need to explain why rank $N^{cd}$ is $5$, i.e. because
% there's a conservation of number of enzyme states?}}

As discussed in \S \ref{sec:pathway-analysis}, two positive pathways
were identified both by bond graph pathway analysis and using
metatool%
\footnote{The Octave~\citep{EatBatHau15} version of metatool from
  \url{http://pinguin.biologie.uni-jena.de/bioinformatik/networks/metatool/metatool5.0/metatool5.0.html}
was used.}
. The corresponding PPM  $\Kp$ is:
\begin{equation}\label{eq:K_P_Hill}
  % \Kp^T =
  % \begin{pmatrix}
  %   1&1&1&1&1&1&0\\
  %   1&1&0&1&0&0&1
  % \end{pmatrix}^T
  \Kp^T =
  \begin{pmatrix}
    1&1&1&1&1&1&0\\
    1&0&0&0&1&1&1
  \end{pmatrix}
\end{equation}
%%\TBD{\added{Is this just one of several possible PPM  $\Kp$?}}
Apart from the ordering of the two columns, the PPM for this system is
unique.
These two columns correspond to the path involving the six reactions:
\{e em lem es esm lesm\} and the upper loop involving the four
reactions: \{e em es slip\}. These are the only positive pathways for
this system. For convenience, the first (outer) pathway will be named
\emph{loop} and the second pathway (involving the ``slip'' reaction)
will be named \emph{slip}; this nomenclature is indicated in Figure
\ref{subfig:Hill_diagram}.
\begin{figure}[htbp]
  \centering
  \SubFig{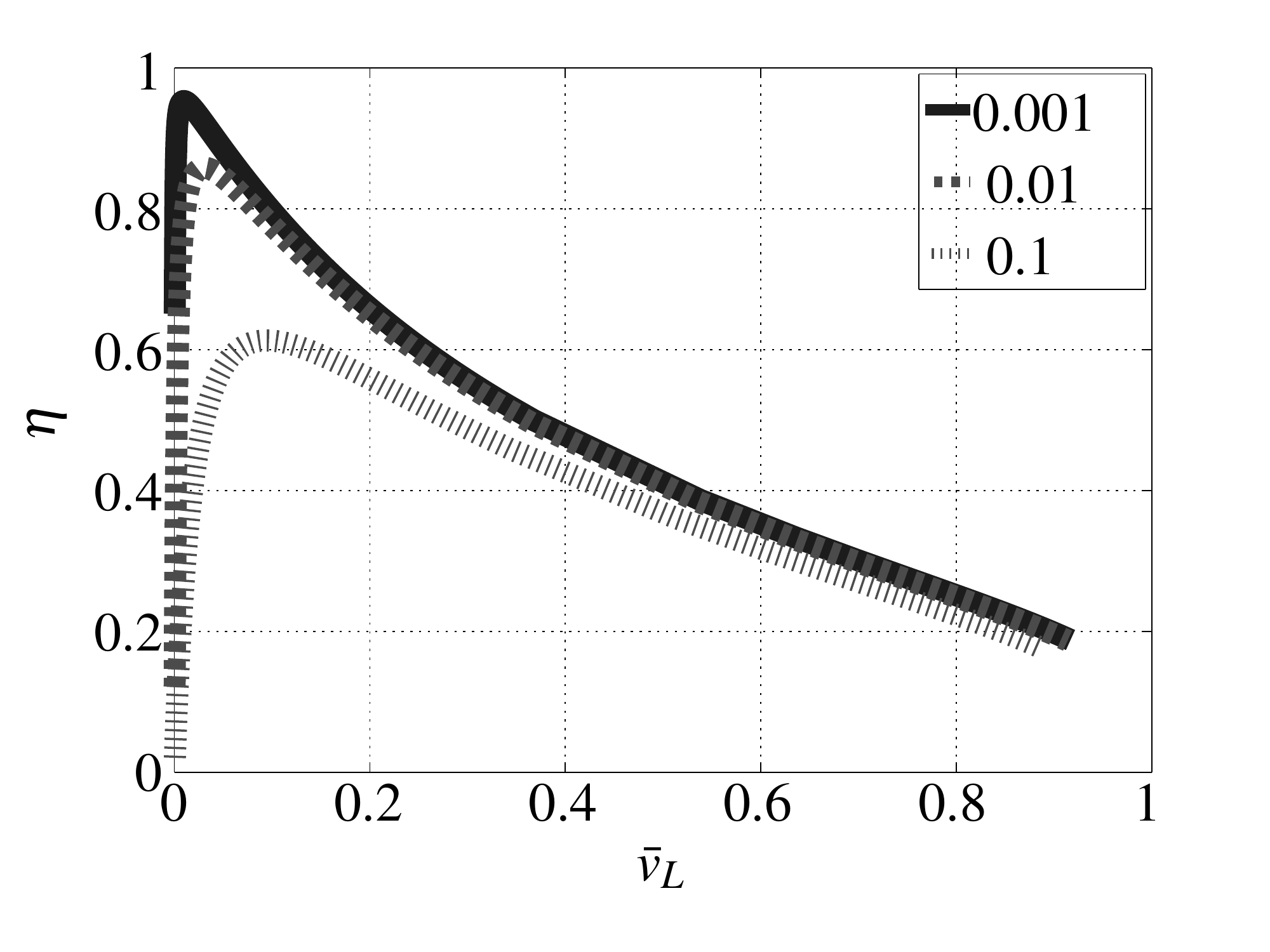}{Efficiency $\eta$ for different $\kappa_{slip}$}{0.45}
  \SubFig{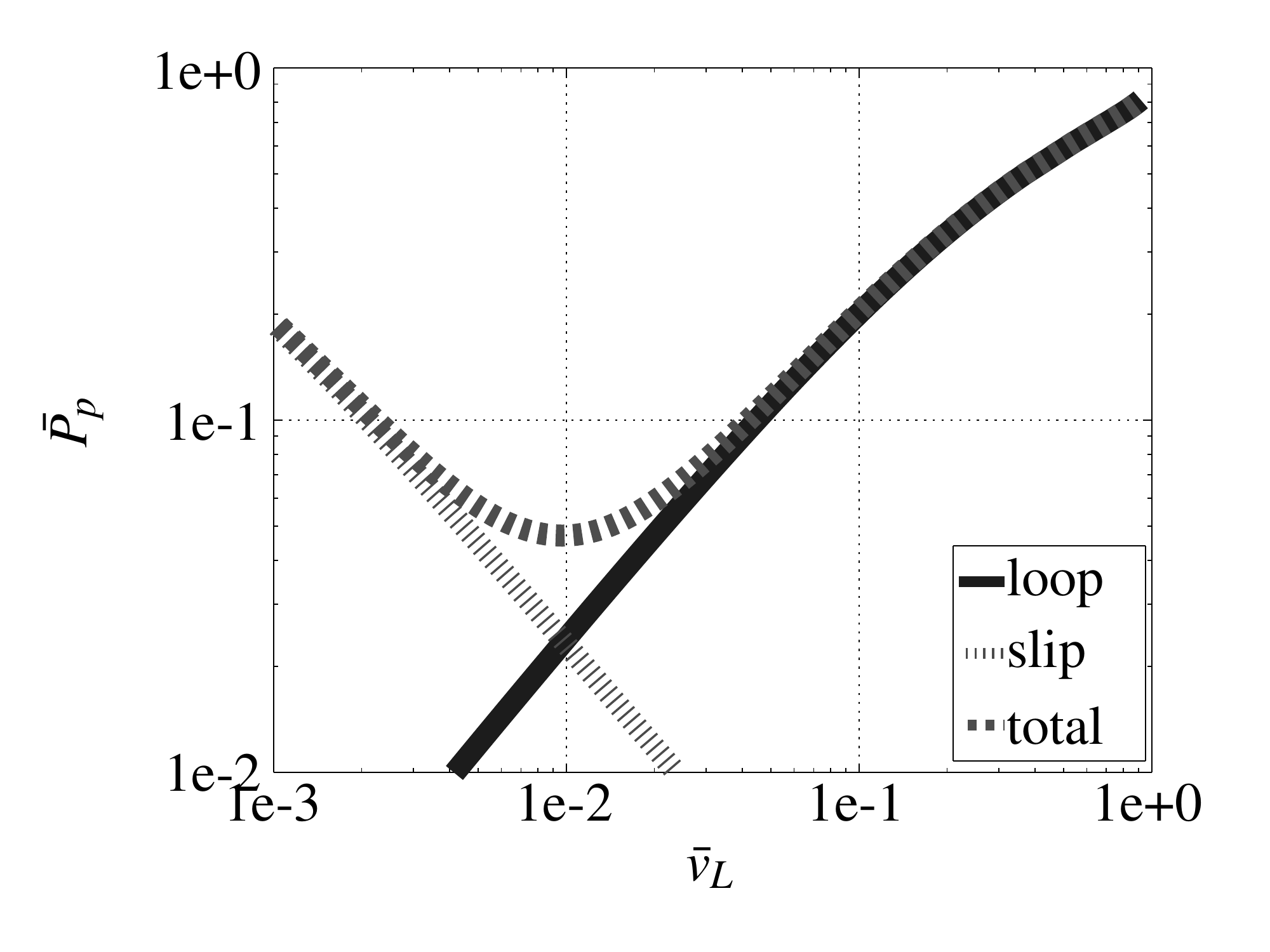}{Normalised dissipation $\Ppb$: $\kappa_{slip} = 10^{-3}$}{0.45}
  \SubFig{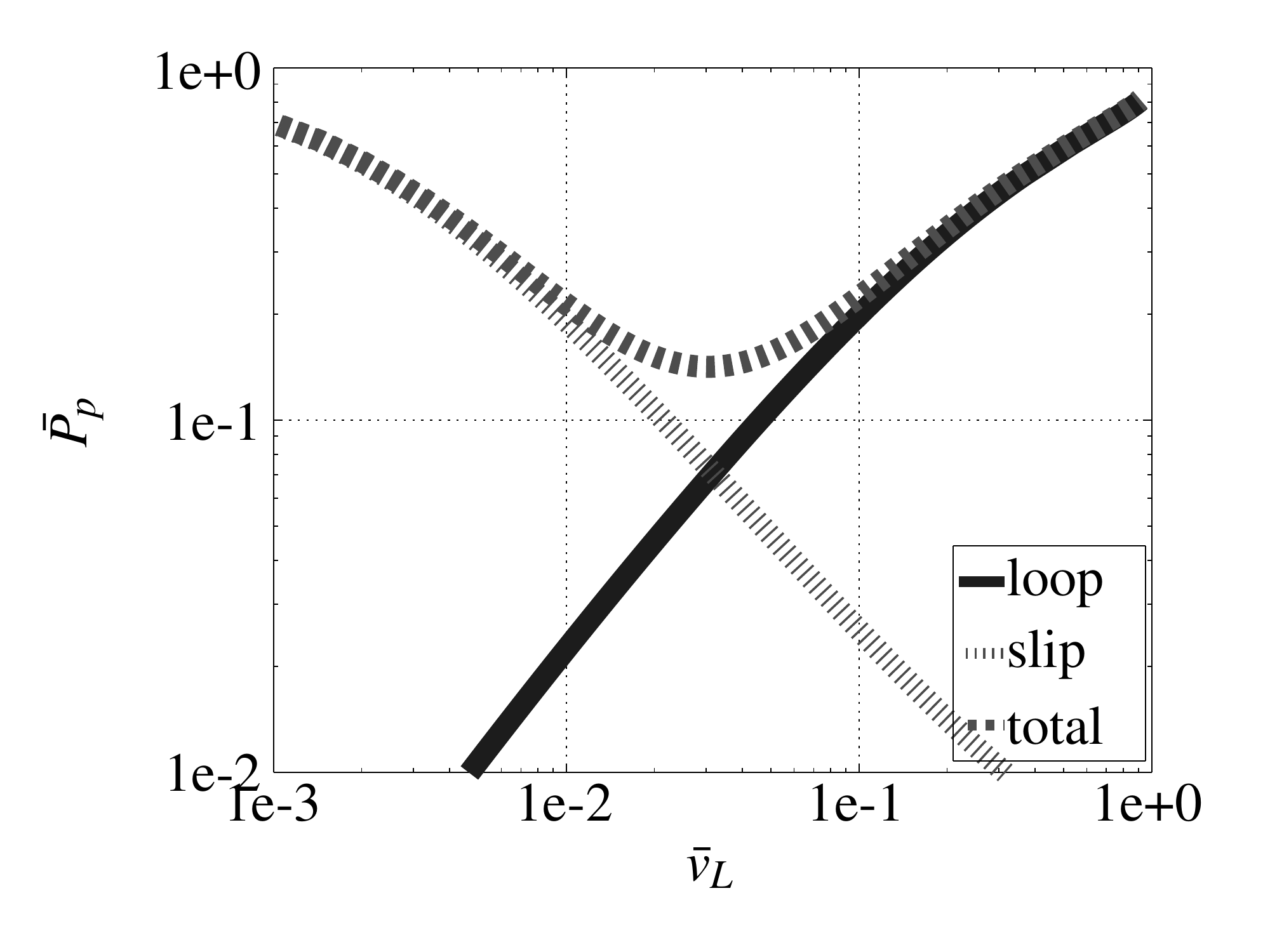}{Normalised dissipation $\Ppb$: $\kappa_{slip} = 10^{-2}$}{0.45}
  \SubFig{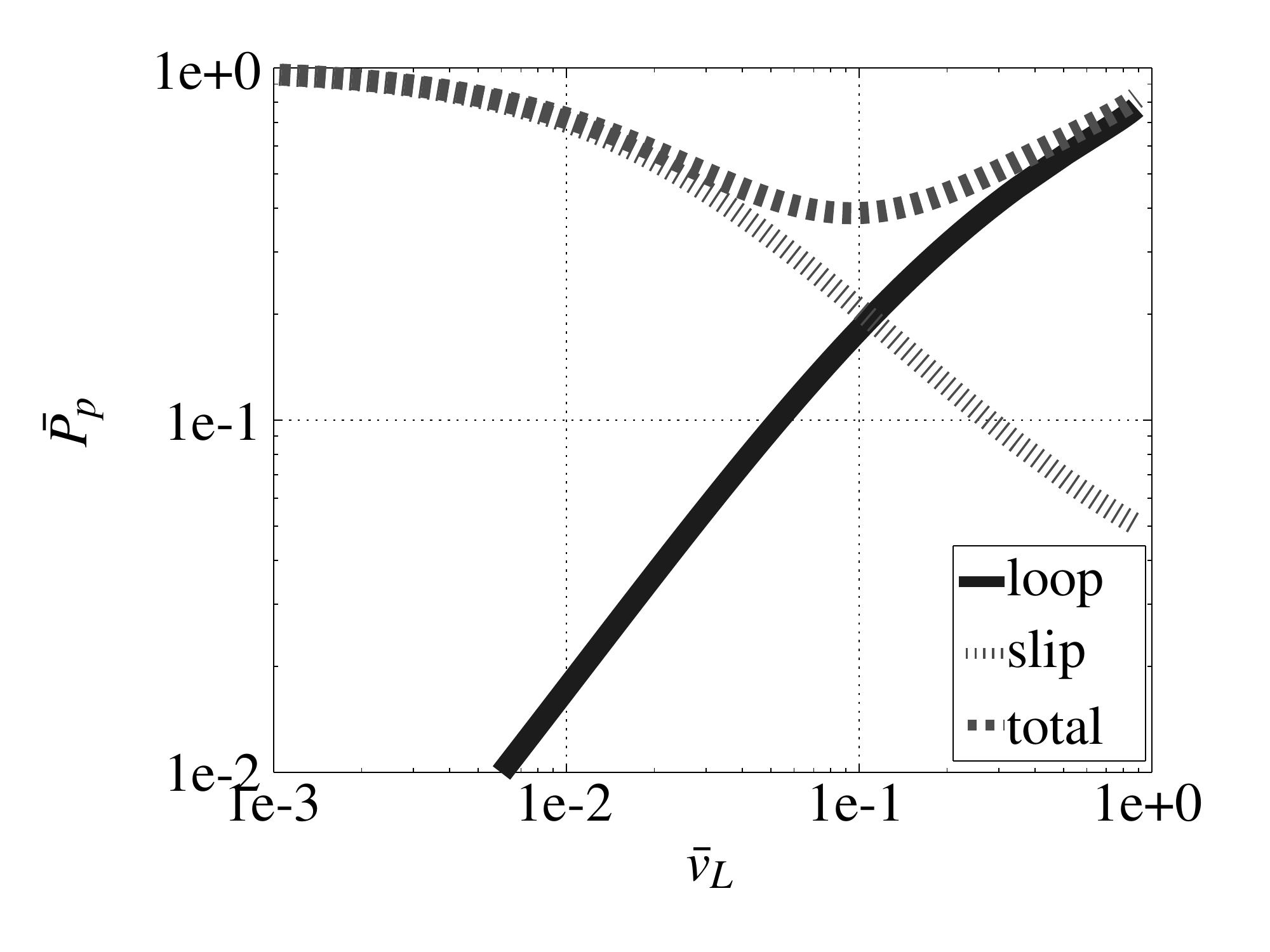}{Normalised dissipation $\Ppb$: $\kappa_{slip} = 10^{-1}$}{0.45}
  %%\SubFig{Hill_slip_0}{$\kappa_{slip} = 1$}{0.45}
  \caption{Free energy transduction and biomolecular cycles:
    Energy-based analysis.  (a) The free energy transduction
    efficiency $\eta$ is plotted against normalised flow $\bar{v}_L$
    of $L$ for the three values of the slippage coefficient
    $\kappa_{slip}$ used in b)~--~(d).  (b)~--~(d) The normalised pathway free energy
    dissipation $\bar{\Pp}$ is plotted agains normalised flow
    $\bar{\VV}_L$ of $\Sp{L}$ for the two pathways \emph{loop} and
    \emph{slip} marked on Figure \ref{subfig:Hill_diagram} together
    with the total normalised free energy dissipation.}
\label{fig:hill-energy}
\end{figure}

\begin{figure}[htbp]
  \centering
  \SubFig{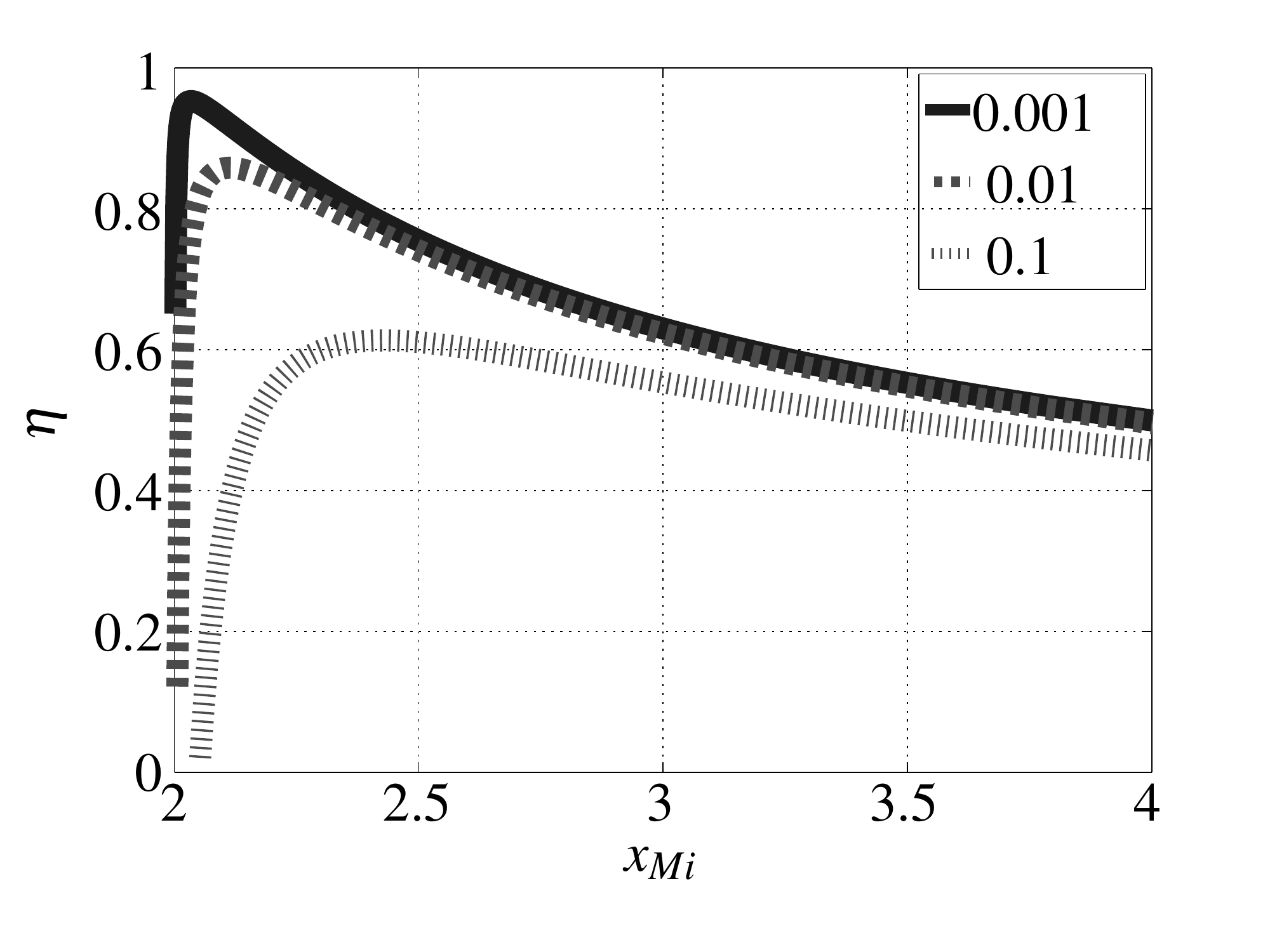}{Efficiency $\eta$ for different $\kappa_{slip}$}{0.45}
  \SubFig{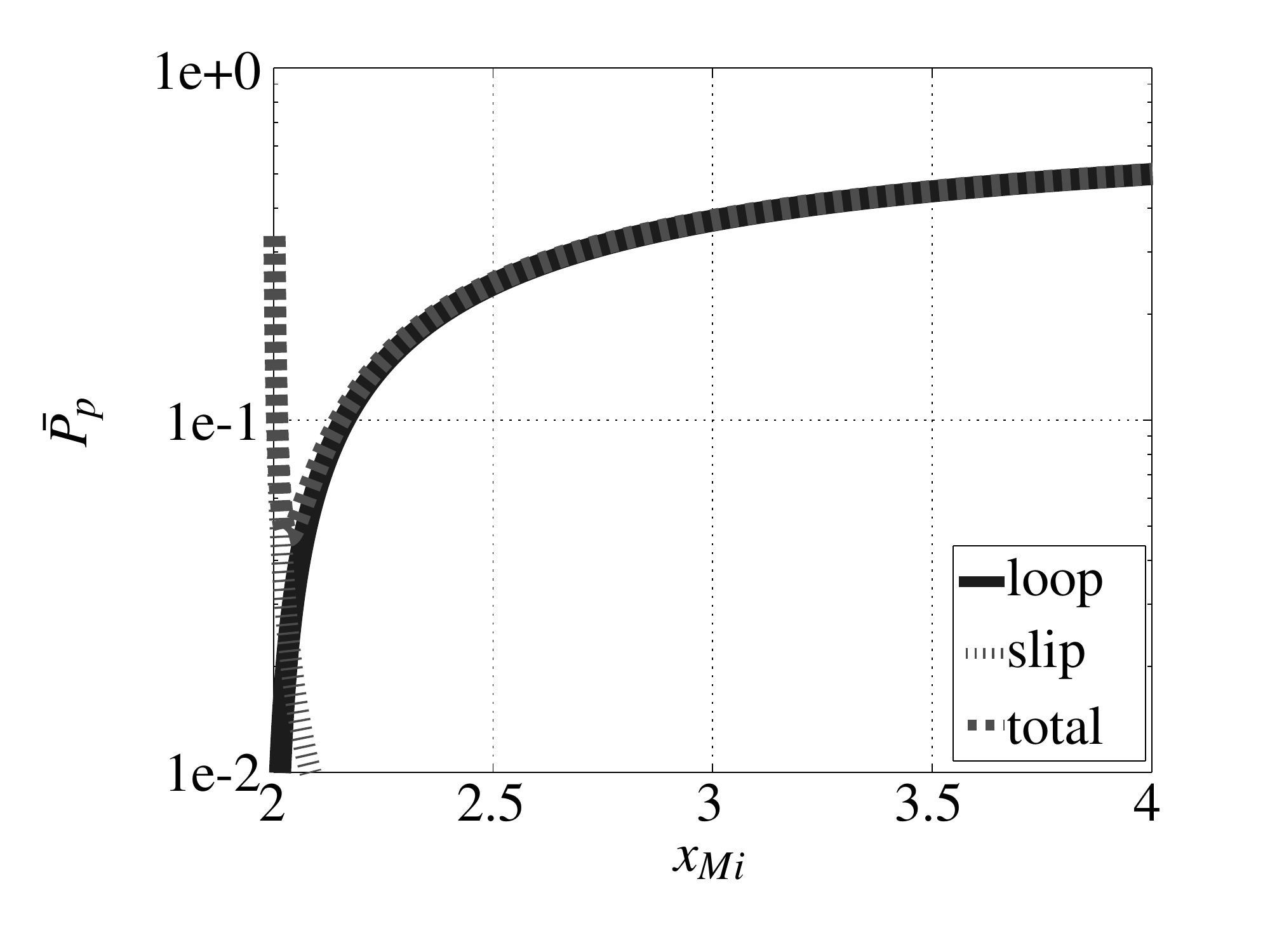}{Normalised dissipation $\Ppb$: $\kappa_{slip} = 10^{-3}$}{0.45}
  \SubFig{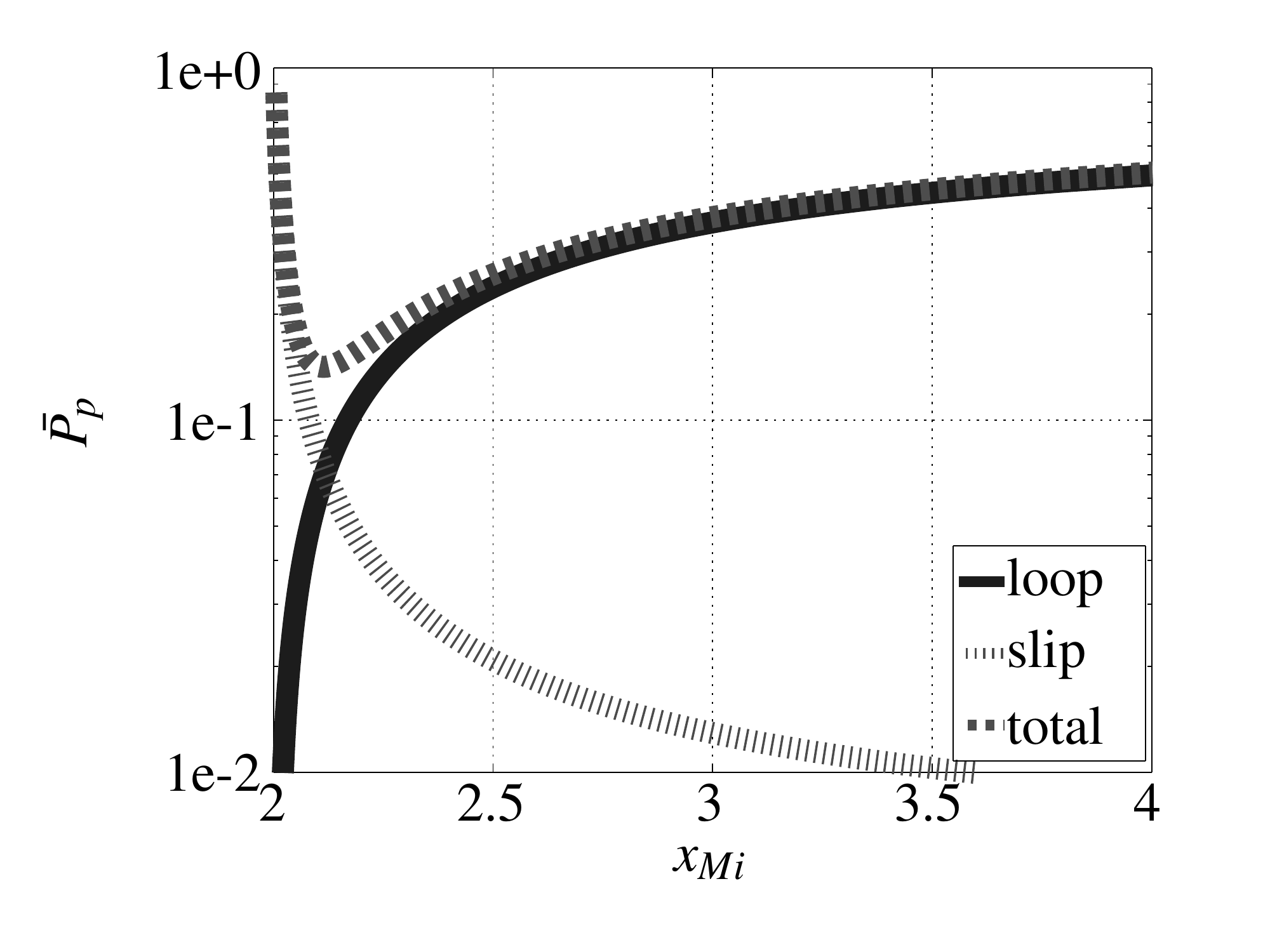}{Normalised dissipation $\Ppb$: $\kappa_{slip} = 10^{-2}$}{0.45}
  \SubFig{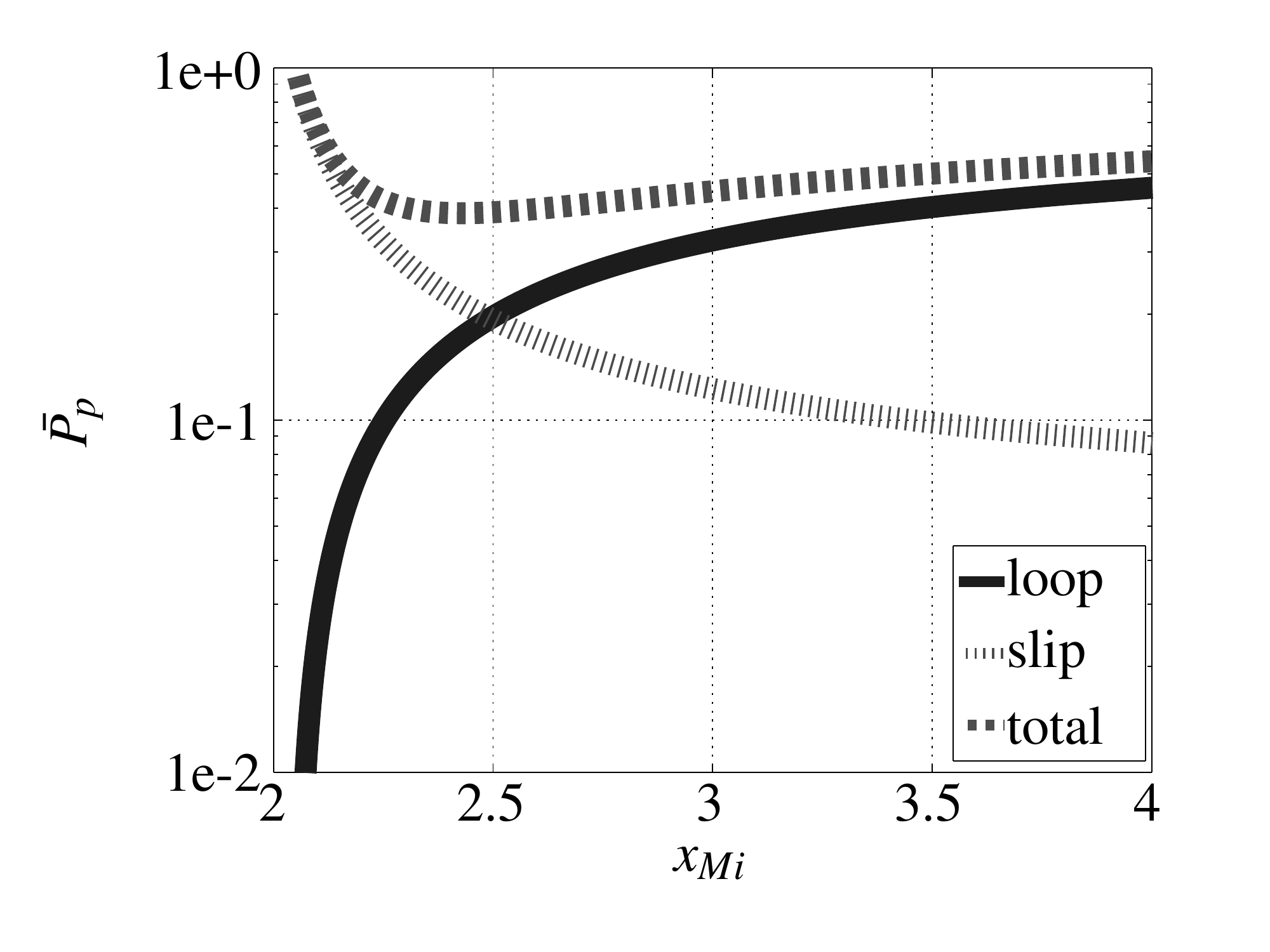}{Normalised dissipation $\Ppb$: $\kappa_{slip} = 10^{-1}$}{0.45}
  %%\SubFig{Hill_slip_0}{$\kappa_{slip} = 1$}{0.45}
  \caption{Free energy transduction and biomolecular cycles:
    Energy-based analysis. Figure \ref{fig:hill-energy} is replotted
    with $x_{Mi}$ replacing $\bar{\VV}_L$}
\label{fig:hill-energy_Mi}
\end{figure}

For the purposes of illustration the thermodynamic constants of the
ten species are taken to be unity, the total amount of $\Sp{E}$, $\Sp{EM}$,
$\Sp{LEM}$, $\Sp{E^\star}$, $\Sp{E^\star M}$ and $\Sp{LE^\star M}$ is taken as $\Sp{E_{tot}=10}$,
the amount of $\Sp{M}_o$ and $\Sp{L}_i$ is taken as unity, the amount
of $\Sp{L}_o$ as two and $\Sp{M}_i$ is variable. The rate constant of
the slippage reaction is varied in the sequel, and that of the other
reactions is $10$.

The system was simulated with a slow change of the amount of $\Sp{M}_i$
from 2 to 100 such that the system was effectively in a steady state
for each value of the amount of $\Sp{M}_i$; these values were then refined
using a steady-state finder. The conserved moiety was automatically
accounted for using the method of \citet[\S3(c)]{GawCra14}.  The
results were checked using explicit expressions for the steady state
values using the software of \citet{QiDasHan09} which is based on the
method of King \& Altman. The system equations are given in the
Supplementary Material.

Figure \ref{fig:hill_SS_path_pos} shows the two pathway flows $v^P_1$
and $v^P_2$ against the amount $x_{Mi}$ of $Mi$ for four values of
$\kappa_{slip}$. As derived in the Supplementary Material,
the maximum flow when the $\kappa_{slip}=0$ is given by
$v_{max}=\frac{100}{21} = 4.76$ and this value is also plotted on each graph.

The purpose of this cycle is to use the excess of $M$ to pump $L$
against the concentration gradient. In addition to viewing the cycle as
a mass transport system, it may also be viewed as a system for
transducing free energy from $M$ to $L$. From this point of view, it is
interesting to investigate the (steady-state) \emph{efficiency} $\eta$ of the energy
transduction \citep[\S1.4]{Hil89}, which we define as:
\begin{xalignat}{3}
  \eta &= \frac{P_L}{P_M}&
\text{where } P_L &= \lb \mu_{Lo}-\mu_{Li}\rb v_L&
\text{and } P_M &= \lb \mu_{Mi}-\mu_{Mo}\rb v_M\label{eq:P_M}
\end{xalignat}
The simulated $\eta$ is plotted against normalised flow
$\bar{v}_L = \frac{v_{Lo}}{v_{max}}$ in Figure \ref{subfig:Hill_efficiency}
for the four values of $\kappa_{slip}$ and reveals three features:
\begin{enumerate}
\item with negligible slip, the maximum efficiency occurs at low flow
  rates.
\item the efficiency decreases as $\kappa_{slip}$ increases
\item the flow rate corresponding to maximum efficiency increases
  with $\kappa_{slip}$.
\end{enumerate}
To further invesigate the source of the inefficiency, the pathway
dissipation $\Pp$ was computed and normalised to give
$\Ppb = \frac{Pp}{P_M}$. This is plotted for each pathway, together
with the total normalised dissipation, against the normalised flow
$\bar{v}_L$ of $\Sp{L}$ in Figures
\ref{subfig:Hill_slip_3}~--~\ref{subfig:Hill_slip_1} for three values
of $\kappa_{slip}$. As can be seen, the main loop pathway dissipation
\emph{increases} with flow of $\Sp{L}$ with all of the free energy
associated with $\Sp{M}$ dissipated at the maximum flow rate. In
contrast, the slippage pathway dissipation \emph{decreases} with flow
of $\Sp{L}$ with all of the free energy associated with $\Sp{M}$
dissipated at zero flow rate. The combined normalised dissipation thus
has a minimum (i.e. energy transduction is most efficient) at an
intermediate flow rate dependant on $\kappa_{slip}$.

% Figure \ref{fig:hill_SS_path_pos} shows how the pathway flows $v_P$
% depend in a non-linear fashion on the concentration $x_{Mi}$ of
% $\Sp{M}_i$ in the positive flow regime. 
Figure \ref{fig:hill-energy_Mi} is similar to Figure
\ref{fig:hill-energy} except that the efficiency $\eta$ and normalised
energy flows $\Ppb$ are plotted against  $\Sp{M}_i$ in the positive flow
regime. The minima occur at values of $\Sp{M}_i$ towards the lower end
of the positive flow regime. This reflects the fact that, as shown in
Figure \ref{fig:hill_SS_path_pos}, large changes in flow $v_P$
correspond to small changes in concentration $x_{Mi}$ towards the lower
end of the positive flow regime; for larger values of  $x_{Mi}$, the
effect of changes in  $x_{Mi}$ on $v_P$ is smaller.

\begin{figure}[htbp]
  \centering
  \SubFig{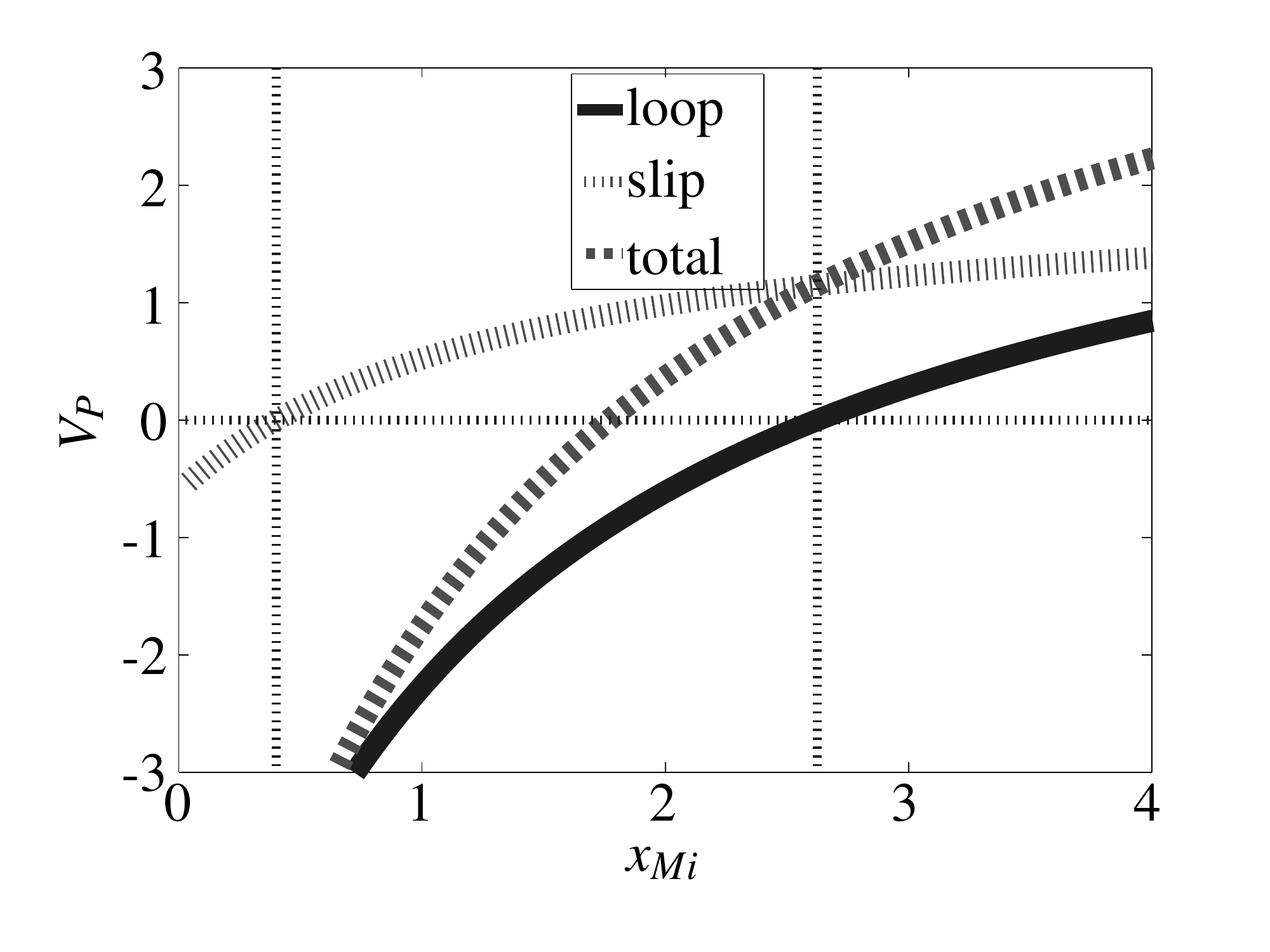}{Mass flows}{0.45}
  \SubFig{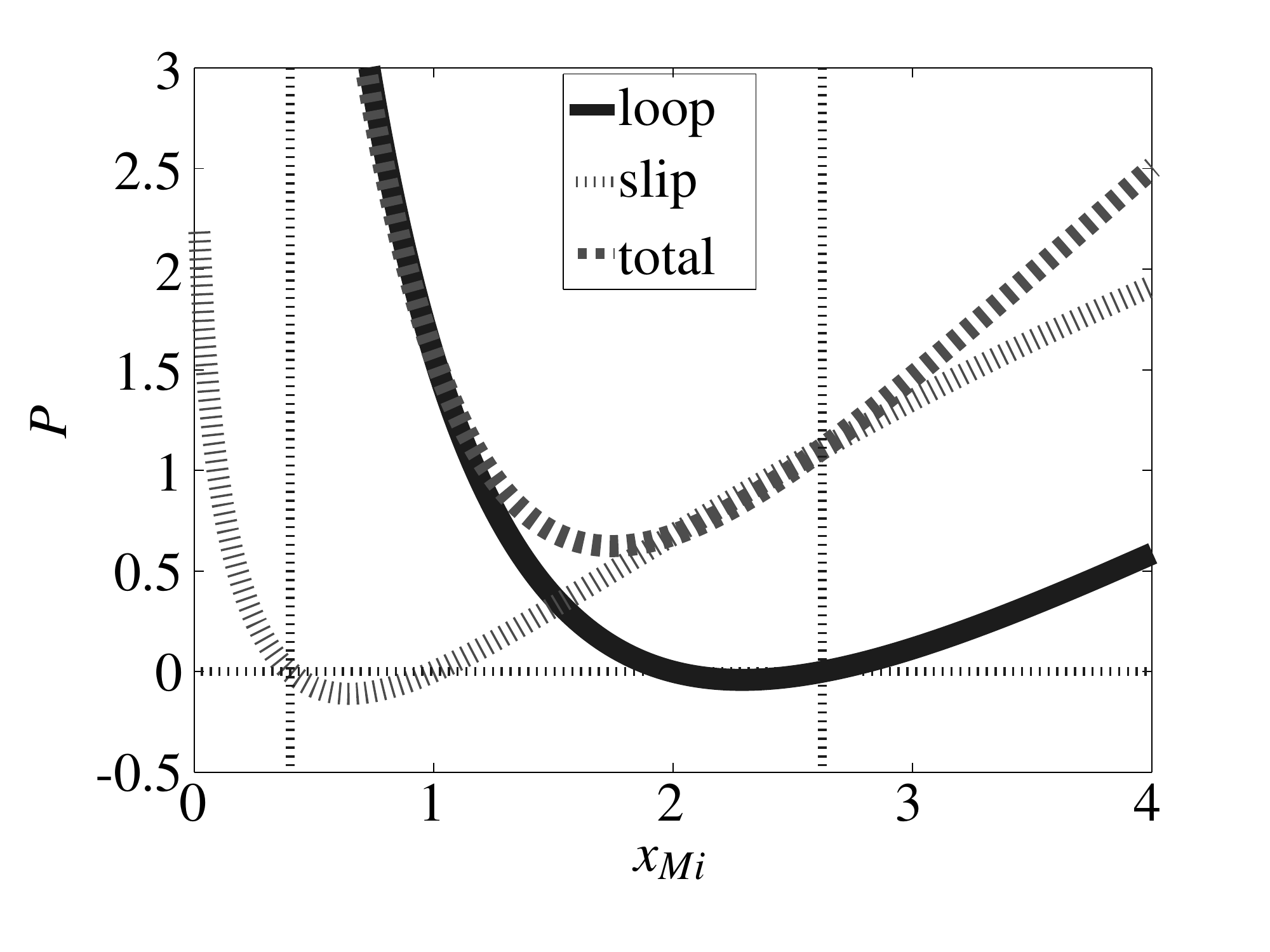}{Energy flows}{0.45}
  \caption{Mass and Energy flows in the crossover region. For
    sufficiently small $\Sp{M}_i$, the cycle no longer acts as a
    transporter as the flows in the two pathways labelled \emph{loop}
    and \emph{slip} in Figure \ref{subfig:Hill_diagram} become
    negative. This figure examines the crossover regime (delineated by
    the two vertical lines) where one pathway flow is positive and the
    other negative -- with a large value ($\kappa_{slip} = 1$) of
    slippage to exaggerate the effect. Notice the negative pathway
    dissipations when the sign of the pathway flows is different -- this
    is due to the intersection of the two pathways leading to a
    non-diagonal $\PW$.}
\label{fig:Hill_cross}
\end{figure}

As mentioned in \S~\ref{sec:pathway-analysis}, the reactions
corresponding to the two pathways of this example are not independent.
To examine the consequences of this interaction, Figure
\ref{fig:Hill_cross} focuses on the system behaviour in the crossover
region where the two pathway flows have opposite directions. 
Figure \ref{subfig:Hill_V_P_slip_0} shows the two mass flows which
have opposite sign in the crossover region $0.42<\Sp{M}_i<2.62$
delineated by the two vertical lines.  
Within this region, Figure \ref{subfig:Hill_P_slip_0} shows that the
individual pathway energy dissipations may be negative; of course, as
indicated in Figure \ref{subfig:Hill_P_slip_0}, the total dissipation
remains positive.  This behaviour is a consequence of the intersection
of the two pathways leading to a non-diagonal $\PW$ (Figure
\ref{fig:EBA}).
Furthermore, the sum of the two pathway flows, corresponding to the
flow of $\Sp{M}_i$ and $\Sp{M}_o$, also changes sign thus causing the
associated energy flow $P_M$ \eqref{eq:P_M} to also change sign. Thus
normalising the pathway energy transduction in the crossover region
using Equation \eqref{eq:P_M} is not helpful.
However, the normal operating region of such a biomolecular cycle is
outside the crossover region and hence this situation does not
normally arise.

% \TBD{\added{Fig.  \ref{fig:Hill_cross}. Expain pathway interaction and
%   apparent negative dissipation in the crossover region due to pathway
%   intersection \& non-diagonal $\PW$. }}

\section{Example: The Sodium-Glucose Transport Protein 1 (SGLT1)}\label{sec:exampl-sodi-gluc}
\begin{figure}[htbp]
  \centering
  \SubFig{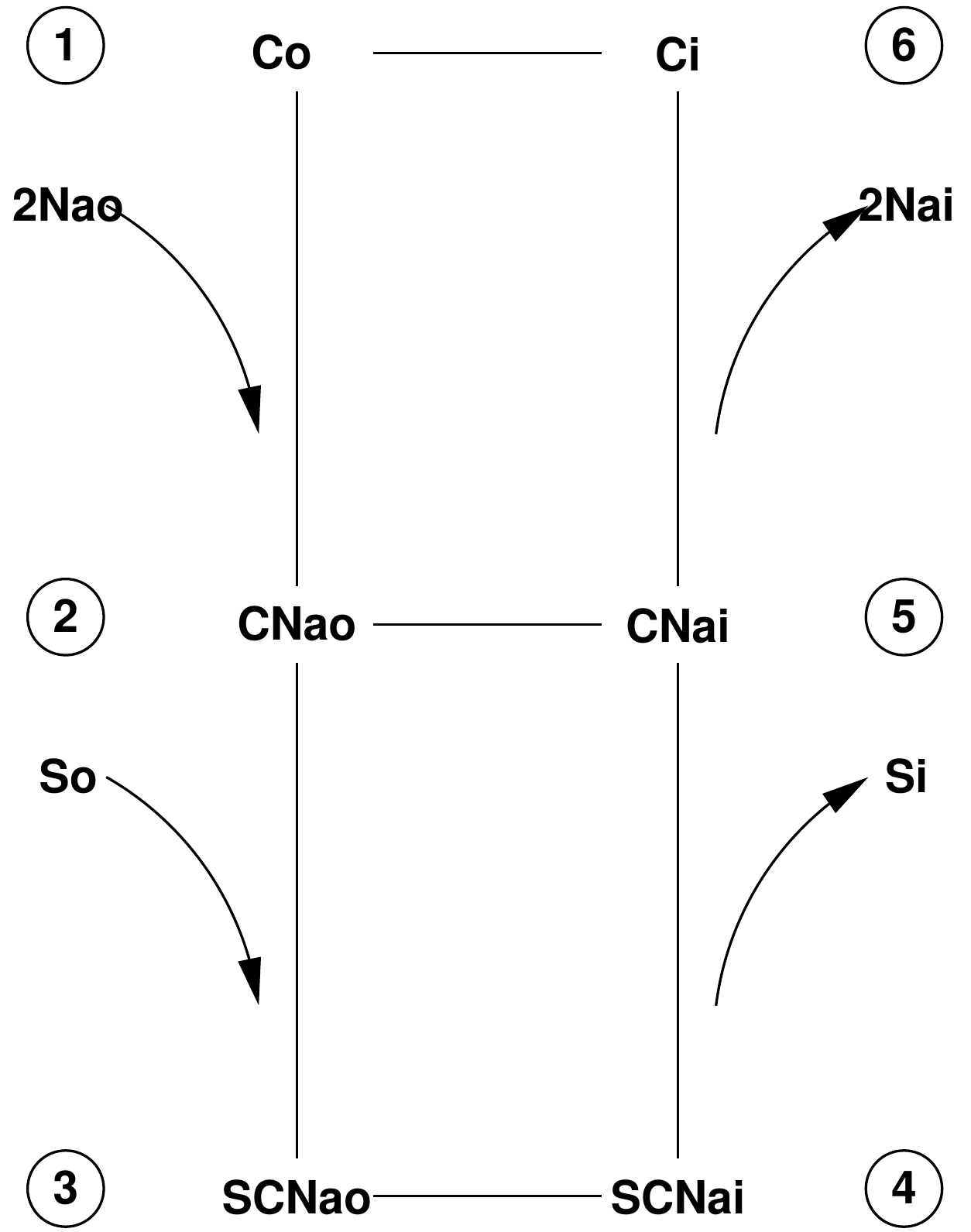}{Diagram}{0.32}
  \SubFig{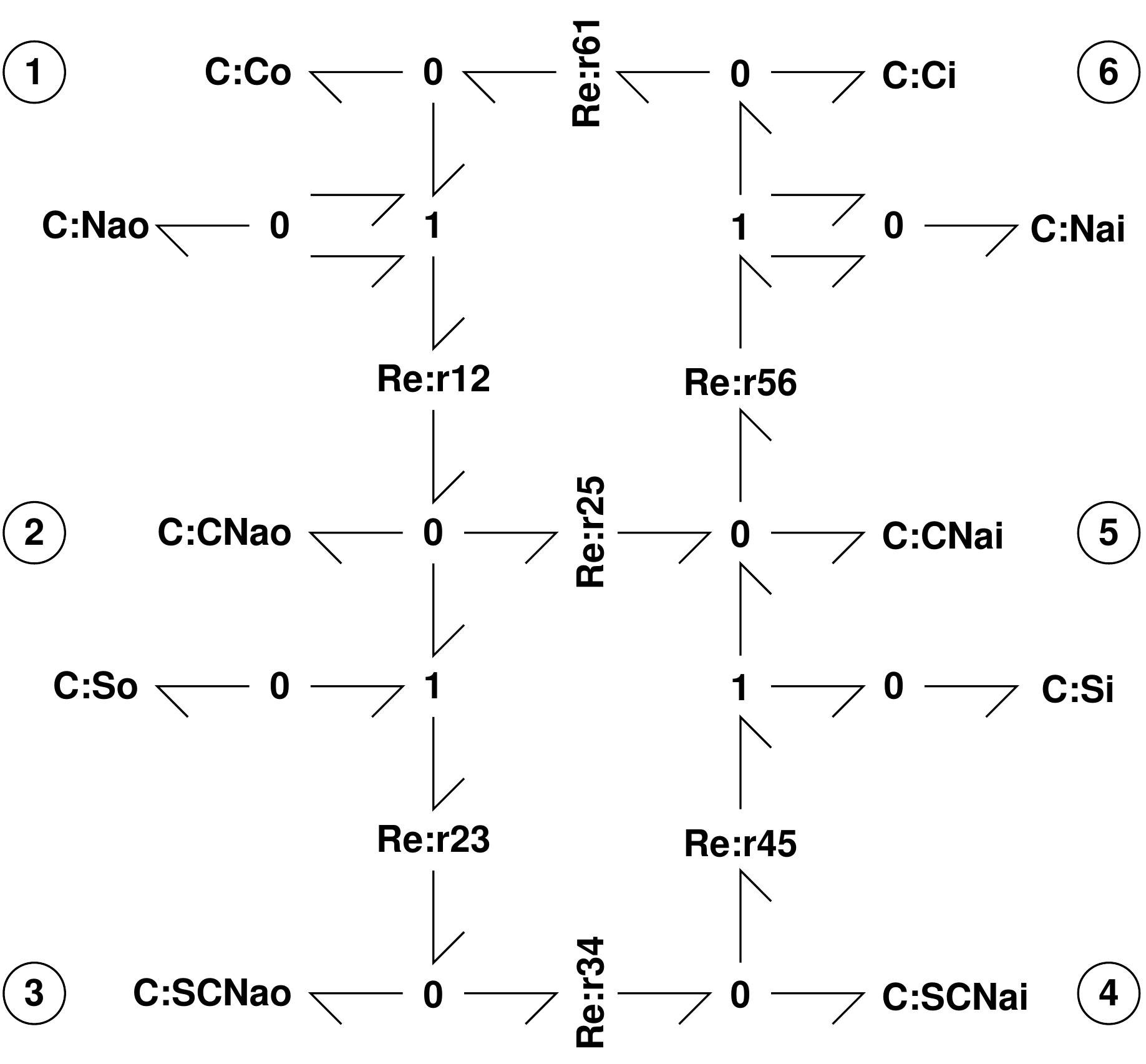}{Bond Graph}{0.45}
  \caption[The Sodium-Glucose Transport Protein 1 (SGLT1)]{The
    Sodium-Glucose Transport Protein 1 (SGLT1). (a) Diagram with the
    six states marked. (b) The corresponding bond graph.}
\label{fig:EskWriLoo05}
\end{figure}
% \begin{table}[htbp]
%   \centering
%   \begin{tabular}{|l||l|l||l|l|}
%     \hline
%     Reaction & $k_f$ & $k_r$ & $K_{eq}=\frac{k_f}{k_r}$ & $\kappa$\\
%     \hline
%     \hline
%     r12 & 80000 & 500 & 160 & 10.1796 \\
%     r23 & 100000 & 20 & 5000 & 202.023 \\
%     r34 & 50 & 50 & 1 & 505.058 \\
%     r45 & 800 & 12190 & 0.0656276 & 8080.93 \\
%     r56 & 10 & 4500 & 0.00222222 & 67.1184 \\
%     r61 & 3 & 350 & 0.00857143 & 8.67804 \\
%     r25 & 0.3 & 0.00091 & 329.67 & 0.00610777 \\
%     \hline
%   \end{tabular}
%   \caption{Reaction Parameters}\label{tab:parameters}
% \end{table}

% \begin{table}[htbp]
%   \centering
%   \begin{tabular}{|l||l|}
% \hline
% Species & $K$\\
% \hline
% \hline
% So & 10.0777 \\
% Si & 10.1247 \\
% Nao & 13.9591 \\
% Nai & 13.9263 \\
% Co & 40.3317 \\
% CNao & 49.1178 \\
% SCNao & 0.0989984 \\
% Ci & 0.3457 \\
% CNai & 0.148991 \\
% SCNai & 0.0989984 \\
% \hline
%   \end{tabular}
%   \caption{Species Parameters}\label{tab:species_parameters}
% \end{table}
The Sodium-Glucose Transport Protein 1 (SGLT1) (also known as the
$\Na$/glucose transporter) was studied experimentally by
\citet{ParSupLoo92} and explained by a biophysical model
\citep{ParSupLoo92a}; further experiments and modelling were conducted
by \citet{CheCoaJac95}.  \citet{EskWriLoo05} examined the kinetics of
the reverse mode using similar experiments and analysis to
\citet{ParSupLoo92,ParSupLoo92a} but with reverse transport and
currents.
This example looks at a bond graph based model of SGLT1 based on the
model of \citet{EskWriLoo05}. For simplicity, it is assumed that the
membrane potential is zero and thus there are no electrogenic effects.

The model of \citet[Figure 6B]{EskWriLoo05}, given in diagrammatic form
in Figure \ref{subfig:EskWriLoo05_diagram} is based on the six-state
biomolecular cycle of Figure 2 of \citet{ParSupLoo92a}. When operating
normally, sugar is transported from the outside to the inside of the
membrane driven against a possibly adverse gradient by the
concentration gradient of $\Na$. The diagram of Figure
\ref{subfig:EskWriLoo05_diagram} is similar to the Hill model of
Figure \ref{subfig:Hill_diagram}. Apart from the renaming of
components, and the reversal of inside and outside, the major
difference is that the single driving molecule $Mi$ and $Mo$ replaced by
\emph{two} driving molecules $2\Nao$ and $2\Nai$. 
This change is reflected in the bond graph of
\ref{subfig:EskWriLoo05_abg} by the double bonds.
This also means that the stoichiometric matrix $N$ is the same as that
of Equation \eqref{eq:N_hill} except that the rows corresponding to
$\Nao$ and $\Nai$ are multipled by 2.
However, as $\Nao$ and $\Nai$ are chemostats, this change does not
affect the PPM  $\Kp$ which is still given by \eqref{eq:K_P_Hill}.
The actual matrices are given in the Supplementary Material.
\begin{figure}[htbp]
  \centering
  \SubFig{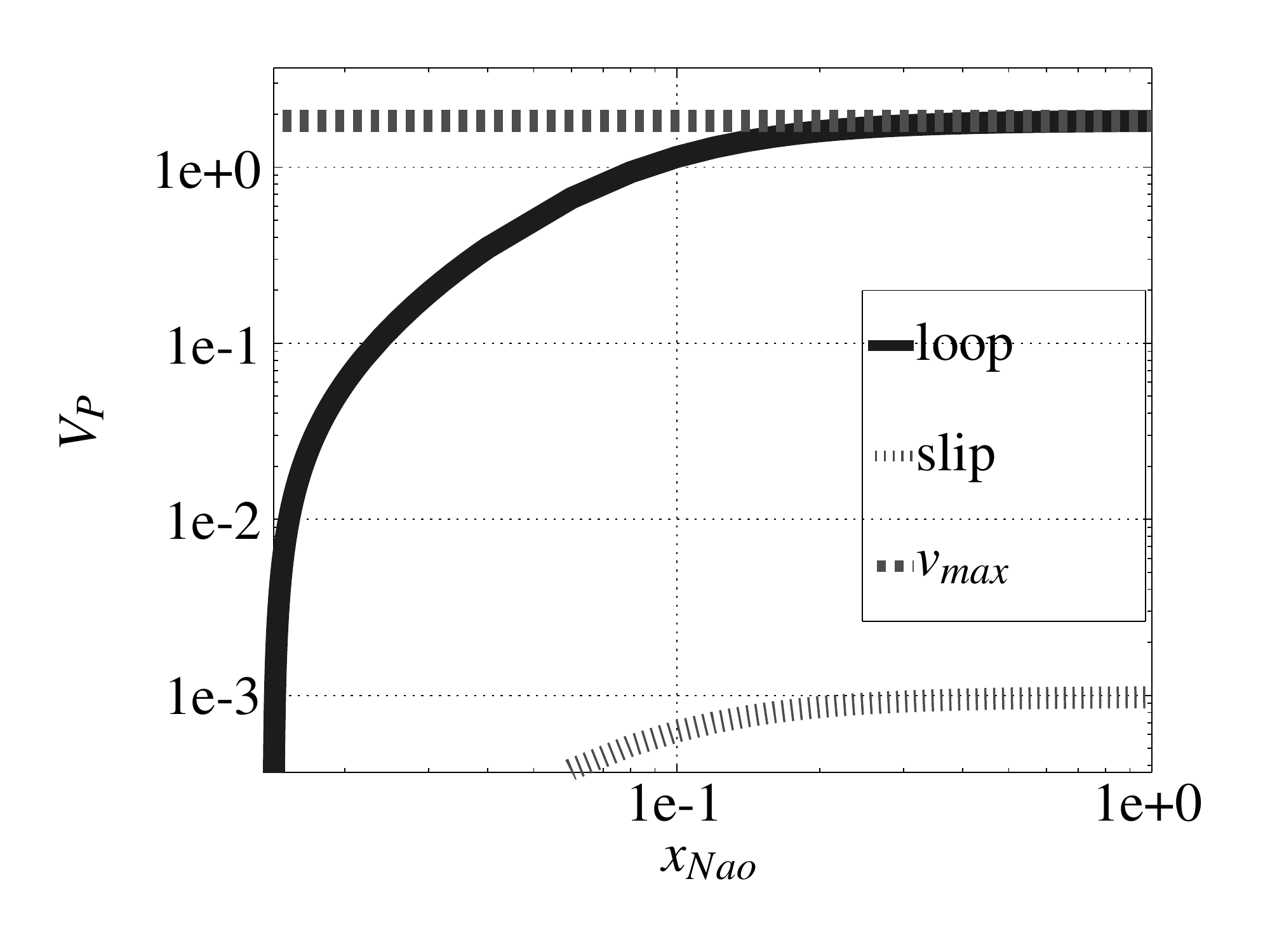}{Pathway flows}{0.45}
  \SubFig{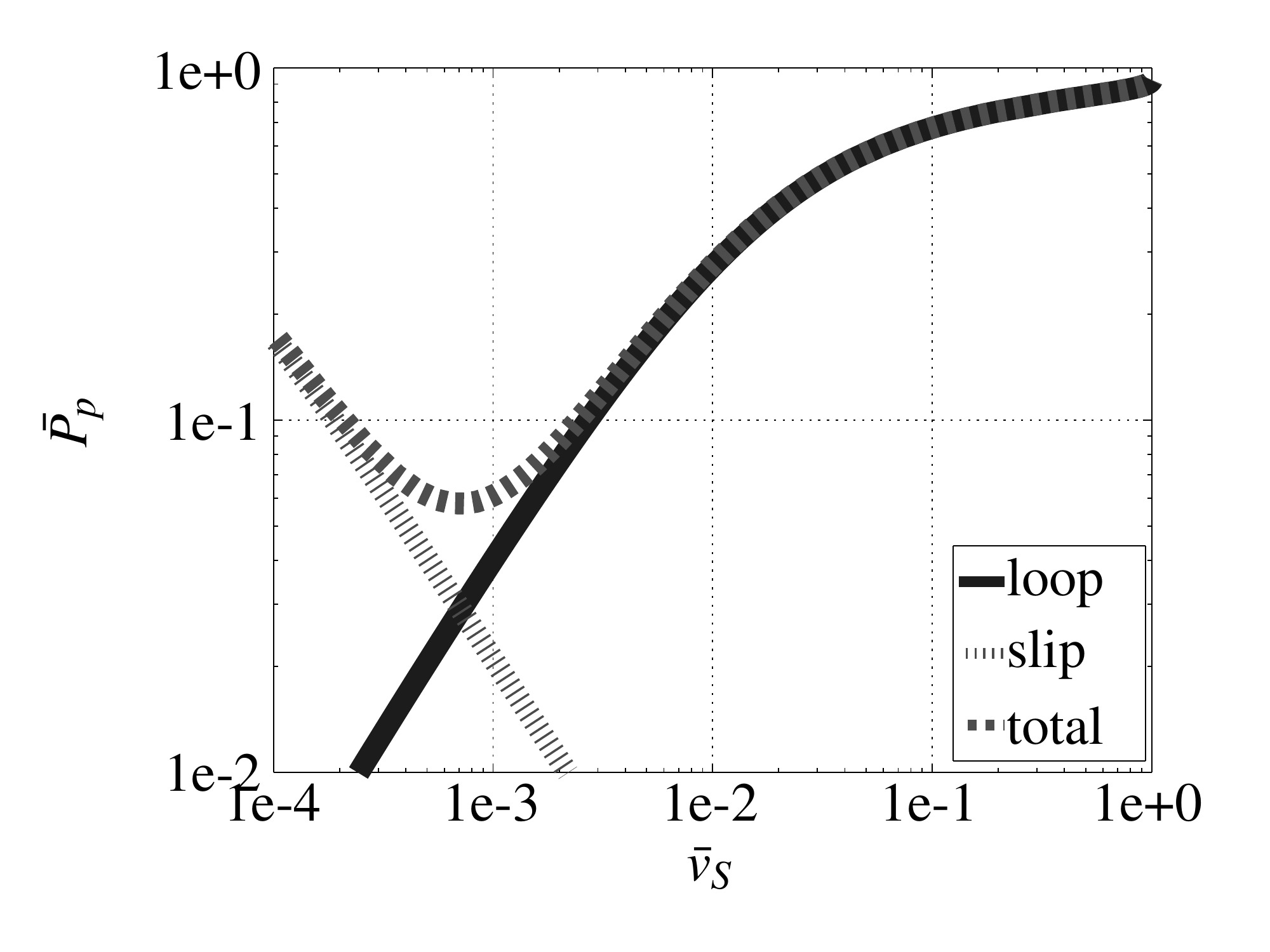}{Pathway dissipation}{0.45}
  \caption[The Sodium-Glucose Transport Protein 1 (SGLT1)]{The
    Sodium-Glucose Transport Protein 1 (SGLT1). (a) Simulated flows
    (normalised) in the two pathways labelled \emph{loop} and
    \emph{slip} in Figure \ref{subfig:Hill_diagram}. (b) The
    normalised pathway dissipations as a function of normalised
    flows.}
\label{fig:EskWriLoo05_sim}
\end{figure}
The seven sets of reaction kinetic parameters are given in
Figure 6B \citet{EskWriLoo05} and listed in the first two columns of
Table 1 of the Supplementary Material and the third column gives the
corresponding equilibrium constants.  The vector of seven equilibrium
constants $K^{eq}$ is converted into the vector of ten thermodynamic
constants $K$ of Table 2 of the Supplementary Material using the formula
of \citet{GawCurCra15}:
\begin{equation}
  \label{eq:K_eq}
  K^{eq}   = \Exp \lb {-N}^T \Ln K \rb
\end{equation}
where $N$ is the $10\times 7$ stoichiometric matrix. 
The corresponding rate constants $\kappa$ are then computed as
discussed by \citet{GawCurCra15} and listed in the final column of
Table 1 of the Supplementary Material.
The system was simulated as described in
\S~\ref{sec:example:-free-energy} and the system equations are given
in the Supplementary Material.

As in Figure \ref{fig:hill_SS_path_pos}, Figure
\ref{subfig:Hill2_SS_slip} shows the two pathway flows and, as in
Figure \ref{fig:hill-energy}, Figure \ref{subfig:Hill2_slip} shows the
corresponding pathway dissipation.  As in
\S~\ref{sec:example:-free-energy}, but using parameters coresponding
to experimental data, the combined normalised dissipation has a
minimum (i.e. energy transduction is most efficient) at an
intermediate flow rate.
\section{Conclusion}
\label{sec:conclusion}

It has been shown that standard methods of mass-flow pathway analysis
can be extended to energy-flow pathway analysis making use of the bond
graph method arising from engineering science.
%
%% Significance of the examples
The method has been applied to a glycolysis example of~\citet[Figure
3.4]{HeiSch96} and the biomolecular cycle of~\citet[Figure
1.2(a)]{Hil89} to enable comparison with standard approaches. 

The analysis of the biomolecular transporter cycle was shown to apply
to a model of the Sodium-Glucose Transport Protein 1 (SGLT1) based on
the experimentally-determined parameters of \citet[Figure
6B]{EskWriLoo05}.  Intriguingly, it was found that the rate of energy
dissipation has a minimum value at a particular normalised flow rate
which in turn corresponds to a particular driving concentration.
This minimum is due to the interaction of a number of factors
including the system parameters, the presence of two interacting
pathways and the concentration of $\Na$ (or $\Sp{M}$ in the case of
\S~\ref{sec:example:-free-energy}) needed to generate the transporter
flow. It would be interesting to compare this theoretical flow rate
to that found in nature.

%% Generality
The bond graph approach can be used to decompose complex systems into
computational modules \citep{GawCurCra15,GawCra16}. Combining such
modularity with energy-based pathway analysis approach of this paper
would provide an approach to analysing and understanding energy flows
in complex biomolecular systems for example those within the Physiome
Project~\citep{Hun16}. This is the subject of current research.

%% Electrogenic.
Although this paper is restricted to flows of chemical energy, the
bond graph approach enables models to be built across multiple energy
domains including chemoelectrical transduction
\citep{Kar90,GawSeiKam15X}. Hence the pathway approach can be equally
well applied to systems with electrogenic features such as excitable
membranes \citep{Hill01} and the mitochondrial electron transport
chain \citep{NicFer13}.

%% Energy and evolution
The effective use of energy is an important determinant of evolution
\citep{Lot22,SouThiLan13,PasProSut13,MarSouLan14}. Therefore the
energy-based pathway analysis of this paper is potentially relevant to
investigating why living systems have evolved as they have. For
example, do real SGLT1 transporters operate near the point of minimal
energy dissipation?

%% Energy and disease.
The supply of energy is essential to life and disruption of energy
supply has been implicated in many diseases such as cardiac
failure~\citep{Neu07,TraLoiCra15}, Parkinson's
disease~\citep{Bea92,WelClo10,SheCai12,WelClo12,Wel12} and
cancer~\citep{MarGalRod11,MarLopGal14,MasAsg14}. Therefore it seems
natural to apply the energy-based methods of this paper to investigate
such systems. 
%
%% Mitochondria
In particular, mitochondria are important for energy transduction in
living systems and mitochondrial dysfunction is hypothesised to be the
source of ageing \citep[Chapter 14]{AlbJohLew15}, cancer
\citep{GogOrrZhi08,SolSgaBar11} and other diseases \citep{NunSuo12}.
Mathematical models of mitochondria exist already
\citep{Bea05,WuYanVin07,CorAon14,BazBeaVin16} and it is hoped that the
energy-based pathway analysis of this paper will shed further light on
the function and dysfunction of mitochondria. This is the subject of
current research.

\subsection*{Data accessibility}
A virtual reference environment \citep{HurBudCra14} is available for
this paper at \url{http://dx.doi.org/10.5281/zenodo.165180}.
The simulation parameters are listed in the Appendix.
\subsection*{Competing interests}
The authors have no competing interests.
\subsection*{Authors' contribution}
All authors contributed to drafting and revising the paper, and
they affirm that they have approved the final version of the
manuscript.
\subsection*{Acknowledgements}
Peter Gawthrop would like to thank the Melbourne School of Engineering
for its support via a Professorial Fellowship. Both authors thank Dr
Ivo Siekmann for discussions relating to conic spaces and Dr Daniel
Hurley for help with the virtual reference environment.
We would also like to thank the reviewers for their suggestions for
improving the paper and Peter Hunter for pointing out some errors in
the draft ms.
\subsection*{Funding statement}
This research was in part conducted and funded by the Australian
Research Council Centre of Excellence in Convergent Bio-Nano Science
and Technology (project number CE140100036).

%%\bibliography{common}

\appendix
\section{A Short Introduction to Bond Graph  Modelling }
\label{sec:short-intr-bond}
The purpose of this section is to provide the bond graph background
necessary to understand the paper itself. More information is to be
found in references \citep{GawCra14,GawCurCra15,GawCra16,Gaw17}.

%% Variables
Bond graphs unify the modelling of energy within and across multiple
physical domains using the concepts of \emph{effort} and \emph{flow}
variables whose product is \emph{power}. Thus in the electrical domain
the effort variable voltage $V$ with units of  \si{V}  and flow
variable current $i$ with units of \si{A} have the
product $P = {Vi}$ with units of \si{J.s^{-1}}. In the biomolecular domain,
the effort variable is \emph{chemical potential} $\mu$ with units of
\si{J \per mol} and the flow variable is reaction flow rate $v$ with units of
\si{mol \per \sec}; $\mu v$ also has units of \si{J.s^{-1}}.

%% Components
The bond graph \C component is a generic energy \emph{storage}
component corresponding to a \emph{capacitor} in the electrical domain
and the amount of a chemical species in the biomolecular domain. If
$q$ is the time integral of the flow variable, the electrical
capacitor generates voltage $V$ according to the \emph{linear}
relationship: $V = q/C$ where $C$ is the capacitance in Farads and $q$
the charge in Coulombs. In contrast, the chemical species generates
chemical potential $\mu\si{J.mol^{-1}}$ according to the
\emph{nonlinear} relationship: $\mu = RT\ln K x$ where $R$ is the
universal gas constant with units of $\si{J.mol^{-1}.K^{-1}}$, $T$
absolute temperature with units of $\si{K}$, $x=q/q_0$ where $q_0$ is
the reference quantity and and $K$ a species-dependant constant which
is dimensionless and depends on the reference quantity.

The bond graph \R component is a generic energy \emph{dissipation}
component corresponding to a \emph{resistor} in the electrical
domain. In the linear case, it generates current $i$ according to the
\emph{linear} relationship: $i = \kappa\Delta V$ where $\kappa$ is the
conductance and $\Delta V$ is the net voltage across the resistor. The
\R component corresponds a chemical \emph{reaction} in the
biomolecular domain. Assuming mass-action kinetics it generates molar
flow $v$ according to the \emph{nonlinear} relationship:
$v = \kappa (\exp A^f/RT - \exp A^r/RT)$ where $A^f$ (the
\emph{forward affinity}) and $A^r$ (the \emph{reverse affinity}) are
the net chemical potentials (with units of \si{J.mol^{-1}}) of the
reactants and products of the reaction respectively; $\kappa$ is the
reaction \emph{rate constant} with units of \si{mol.s^{-1}}. Because
this relationship treats the forward and reverse affinities
differently, this reaction version of the \R component is given a
special symbol: \Re.

%% Bonds and junctions
Bond graphs represent the flow of energy between components by the
bond symbol {\Large $\rightharpoondown$} where the direction of the
harpoon corresponds to the direction of positive energy flow; this is
a sign convention. The bonds connect components via \emph{junctions}
which transmit, but do not store or dissipate energy. There are two
junctions: the \zero junction where all impinging bonds have the same
\emph{effort} and the \one junction where all impinging bonds have the
same \emph{flow}. The expression for the flows associated with the
\zero junction, and the efforts associated with the \one junction, are
determined by the energy conservation requirement. 

%% Example
\begin{figure}[htbp]
  \centering
  \SubFig{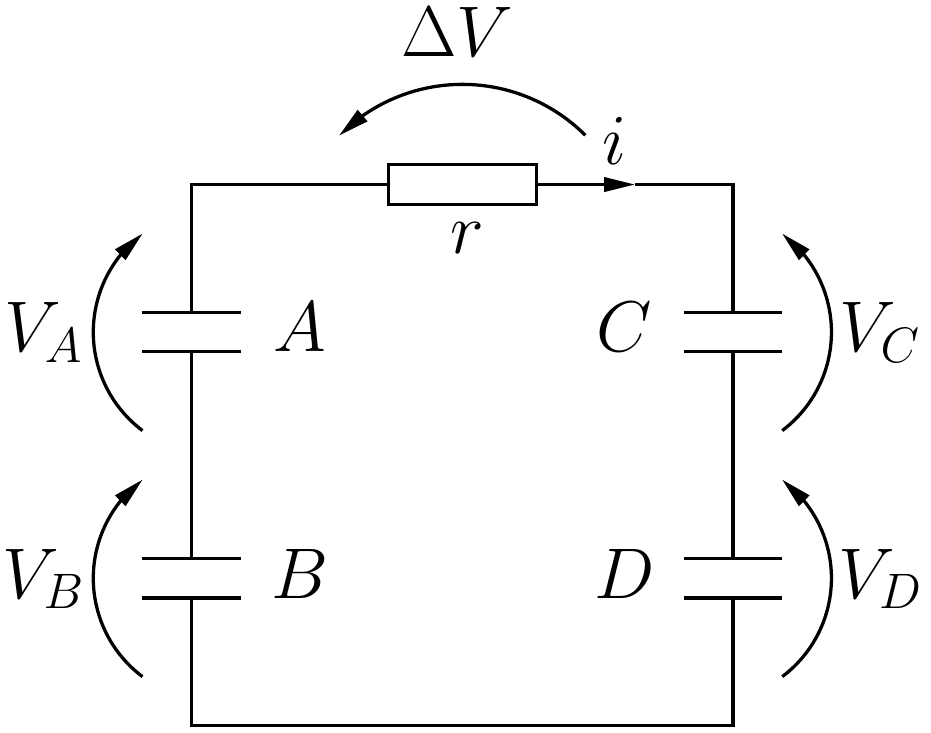}{Schematic}{0.25}
  \SubFig{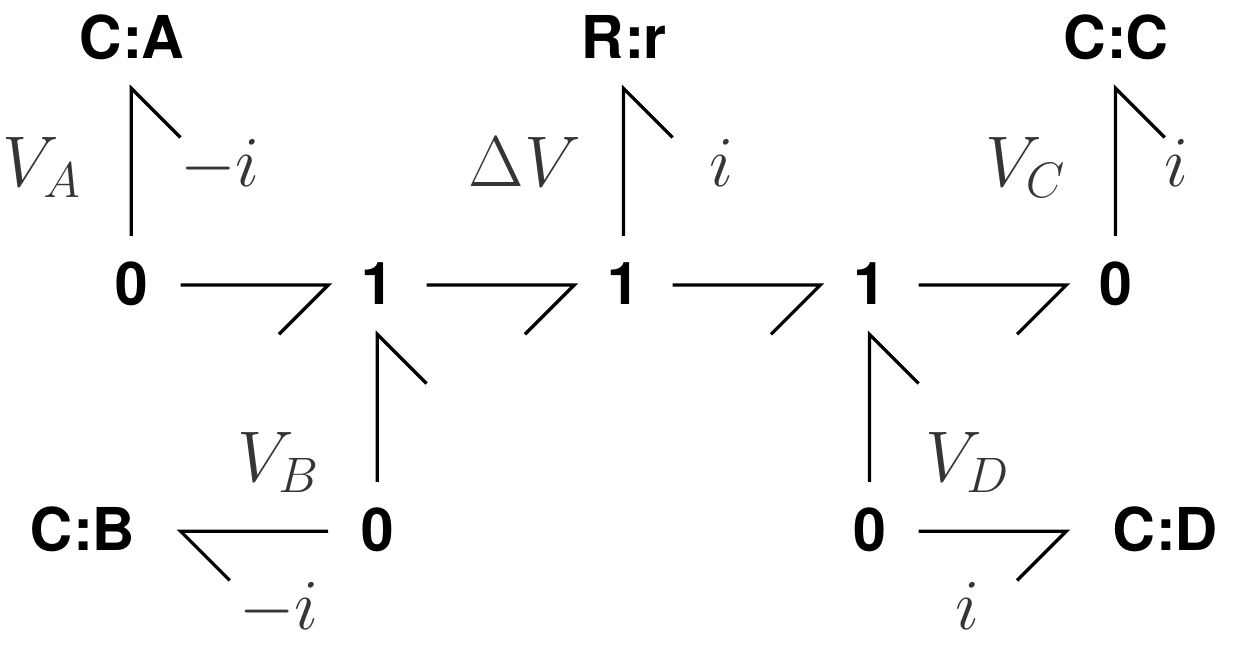}{Bond Graph}{0.45}
  \caption{Modelling Electrical Systems. (a) Electrical schematic
    diagram. (b) Bond graph: $\Delta V = (V_A+V_B) - (V_C+V_D)$.}
  \label{fig:elec}
\end{figure}
Figure \ref{fig:elec} illustrates bond graph  modelling in the
electrical domain. The four capacitors and one resistor are
connected as in the schematic diagram of Figure \ref{subfig:elec}.
In bond graph colon notation, the symbol before the colon indicates the
component type and the symbol after the colon indicates the component
name. Thus the bond graph  of  Figure \ref{subfig:elec_abg} has four
\C components named A--D to represent the four capacitors and a single
\R component named r to represent the resistor. 
The two \zero junctions on the left reverse the sign of the flow
associated with \BC{A} and \BC{B}. The two \zero junctions on the
right are not necessary, but are convenient for connecting to other via bonds
components to form a larger system.
The left \one junction corresponds to $V^f=V_A+V_B$ and the right \one
junction corresponds to $V^r=V_C+V_D$; the centre \one junction
corresponds to $\Delta V=V^f-V^r=(V_A+V_B) - (V_C+V_D)$.
Given the constitutive relations for the \C and \R component, the
dynamical equations describing the system can be automatically
generated from the bond graph  of Figure \ref{subfig:elec_abg}.

%% Example
\begin{figure}[htbp]
  \centering
  \SubFig{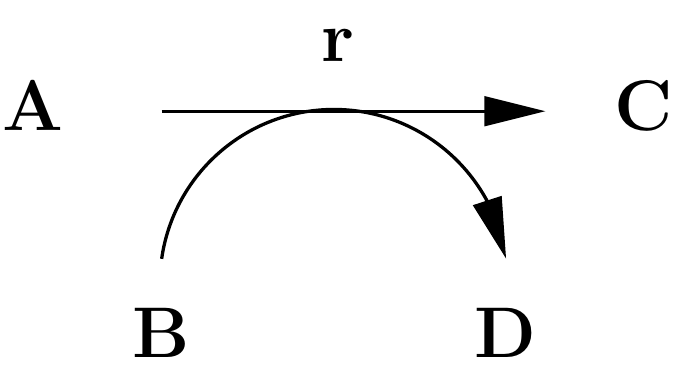}{Schematic}{0.25}
  \SubFig{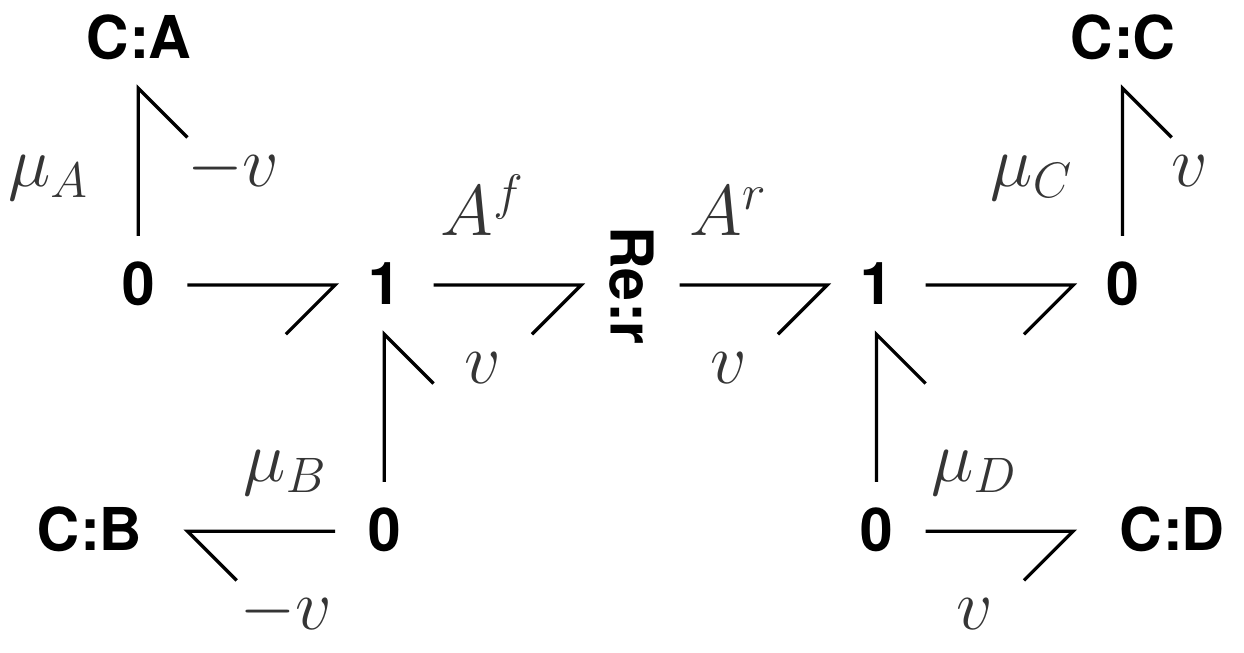}{Bond Graph}{0.45}
  \caption{Modelling Biomolecular Systems: $A+B \reacu{r} C+D$. (a) Biomolecular schematic
    diagram. (b) Bond graph: $A^f=\mu_A+\mu_B$, $A^r=\mu_C+\mu_D$.}
  \label{fig:chem}
\end{figure}
Figure \ref{fig:chem} illustrates bond graph modelling in the chemical
domain; the corresponding reaction is $A+B \reacu{r} C+D$.  In a
similar manner to the electrical system of Figure \ref{fig:elec}, the
bond graph of Figure \ref{subfig:chem_abg} has four \C components
named A--D to represent the four species and a single \Re component
named r to represent the reaction.
The major difference is the use of the \Re component to replace the
\one \& \R combination within the dotted box of Figure
\ref{subfig:elec_abg}; as mentioned previously this is because the
forward and reverse affinities are treated differently.
%
%% Flows
In particular, with reference to the example of Figure \ref{fig:chem}, $A^f=\mu_A+\mu_B$,
$A^r=\mu_C+\mu_D$ which, with reference to Equation \eqref{eq:Afr}
corresponds to 
\begin{xalignat*}{2}
  N^f &=
\begin{pmatrix}
  1 &1 & 0 & 0n
\end{pmatrix}^T&
N^r &=
\begin{pmatrix}
  0 & 0 & 1 & 1
\end{pmatrix}^T
\end{xalignat*}
Using Equations \eqref{eq:V} and \eqref{eq:mu}, it follows that
\begin{align*}
  A^f &= RT \lb \ln K_A x_A + \ln K_B x_B\rb = RT  \ln K_A x_A K_B x_B\\
  A^r &= RT \lb \ln K_C x_C + \ln K_D x_D\rb = RT  \ln K_C x_C K_D x_D
        x_D\\
  \text{and }
  \VV &= \kappa_r \lb e^\frac{\Af}{RT} - e^\frac{\Ar}{RT}\rb = \kappa_r
        \lb K_A x_A K_B x_B - K_C x_C K_D x_D\rb
\end{align*}
This is a form of the mass-action kinetics equation discussed in
the textbooks~\citep{KliLieWie11}.
In particular the equation can be rewritten as:
\begin{align*}
  \VV &= k_f x_A x_B - k_r x_C x_D\\
  \text{where }
  k_f &= \kappa K_A K_B \text{ and } k_r = \kappa K_D K_C
\end{align*}
As mentioned above, the $K$ constants are dimensionless and thus $k_f$
and $k_r$ are dimensionless.  The dimensionless  \emph{equilibrium constant}
$K_{eq}$ is defined as:
\begin{equation*}
  K_{eq} = \frac{k_f}{k_r}
\end{equation*}
As discussed by \citet{GawCurCra15}, $K_{eq}$ and $K_A$, $K_B$, $K_C$
and $K_D$ are related by the general formula \eqref{eq:K_eq}.

%% Conserved Moieties
In both the electrical and biomolecular systems there is only one
flow: the current $i$ in the electrical case and the molar flow $v$ in
the biomolecular case. This flow $f$ is \emph{out} of \BC{A} and
\BC{B} and \emph{into} \BC{C} and \BC{D}. Hence, in the biomolecular
case, $-\dot{x}_A = -\dot{x}_B = \dot{x}_C = \dot{x}_D =v$. This
implies that $x_A + x_C = x_{AC}$, $x_A + x_D = x_{AD}$ and
$x_B + x_C = x_{BC}$ where $x_{AC}$, $x_{AD}$ and $x_{BC}$ are
constant. Each of these three equations represents a \emph{conserved
  moiety}: the total amount of the charge or species involved remains
constant.  The choice of conserved moieties in a given situation is
generally not unique; this is discussed further in
\S~\ref{sec:short-intr-syst}.
% Defining the vector $X=\begin{pmatrix}x_A&x_B&x_C&x_D\end{pmatrix}^T$, the differential
% equations may be rewritten as $\dot{X} = Nv$ where the
% \emph{stoichiometric matrix} $N$ is given by
% $N =
% \begin{pmatrix}
%   -1 &-1 & 1 & 1)
% \end{pmatrix}^T$.
% Examination of the matrix $G$ where $GN=0$ leads to a rigorous
% analysis of conserved moieties. Mathematically $G$ defines the
% \emph{left null space} of the stoichiometric matrix $N$; this is
% discussed further in \S~\ref{sec:short-intr-syst}.

%% Relation to paper
Replacing A,B,C and D by GLC, G6P, ATP and ADP respectively and r by
r4 in Figure \ref{fig:chem} corresponds to the reaction to the left of
the diagram in Figure \ref{fig:HeiSch96}. The apparently redundant
\zero junctions in Figure \ref{fig:chem} are used to provide
connections between this particular reaction and the overall reaction
network of Figure \ref{fig:HeiSch96}.  The reaction diagram of Figure
\ref{subfig:HeiSch96} has the entities ATP and ADP represented more
than once. Although this enhances clarity by removing the need for
intersecting lines on the diagram, it reduces clarity by having each
entity appear in multiple locations. In contrast, the bond graph of
Figure \ref{subfig:HeiSch96_abg} represents ATP and ADP each by a
single \C component (\BG{ATP} and \BG{ADP} respectively) with
appropriate connections to each of the reactions represented by
\BRe{r1}~--~\BRe{r9}.

This idea of representing biomolecular reaction networks by \C
components (representing species) and \Re components (representing
reactions) connected by bonds ({\Large $\rightharpoondown$}) and \zero
and \one junctions is summarised and formalised in Figure
\ref{subfig:Open_bg}.  This representation is summarised in the
caption to Figure \ref{fig:EBA} and is discussed in detail by
\citet{GawCra14,GawCra16}.

Bond graphs provide one foundation for this paper, the other is the
systems biology concept of \emph{pathways} outlined in the next
section.

\section{A Short Introduction to Systems Biology}\label{sec:short-intr-syst}
The purpose of this section is to provide the systems biology
background necessary to understand the paper itself.

%% Stoichiometry
As discussed in the basic textbooks (for example that of
\citet{KliLieWie11}), the \emph{stoichiometric matrix} $N$ is a
fundamental construct in describing and understanding biomolecular
systems. In particular, given $n_X$ chemical species with molar
amounts contained in the column vector $X$ and $n_V$ reactions with
molar flow rates contained in the column vector $V$; the $n_X \times
n_V$ stoichiometric matrix  $N$ relates the rate of change $\dot{X}$
of $X$ to $V$ by Equation \eqref{eq:dX} repeated here:
\begin{xxalignat}{3}
  &&\dot{X} &= N V&&(2.1)
\end{xxalignat}
The elements of $N$ are integers that determine the amount of each
species participating in the reaction. For example, the reaction $A+B
\reacu{r} C+D$ of Figure \ref{fig:chem} can be represented by Equation \eqref{eq:dX} where:
\begin{xalignat*}{3}\label{eq:simple}
  X &=
  \begin{pmatrix}
    x_A&x_B&x_C&x_D
  \end{pmatrix}^T&
 N &=
\begin{pmatrix}
  -1 &-1 & 1 & 1)
\end{pmatrix}^T&
V &= v_r
\end{xalignat*}
where $x_A$ is the molar amount of substance $A$ etc and $v_r$ is the
reaction flow.

%% Null spaces.
As discussed in the basic textbooks (for example that of
\citet{KliLieWie11}), the mathematical concept of a \emph{null space}
of $N$ gives useful information about the fundamental properties of
the biomolecular system described by $N$. These are two null spaces:
the left null space described by the matrix $G$ where $GN = 0$ and the
right null space described by the matrix $\Kp$ where $N\Kp = 0$.
In the case of the open systems discussed in
\S~\ref{sec:energy-based-react}, $N$ is replaced by $\Ncd$.

The significance of $G$ is that pre-multiplying both sides of Equation
\eqref{eq:dX} by $G$ gives
\begin{equation*}
  G\dot{X} = GN V = 0
\end{equation*}
Hence the linear combinations of $X$ implied by the rows of $G$ are
\emph{constant}; in systems biology nomenclature such combinations are
known as \emph{conserved moieties}. In case of the example system, one
possible matrix $G$ is:
\begin{equation*}
  G =
  \begin{pmatrix}
    1&0&1&0\\
    1&0&0&1\\
    0&1&1&0
   \end{pmatrix}
\end{equation*}
It can be verified that $GN=0$ and the corresponding three conserved
moieties are
\begin{xalignat*}{3}
  x_A &+ x_C& 
  x_A &+ x_D& 
  x_B &+ x_C
\end{xalignat*}

The significance of  $\Kp$ is that if the elements of $V$ are not
independent but rather defined by $V = \Kp v$ then Equation
\eqref{eq:dX}  gives
\begin{equation*}
  \dot{X} = N \Kp v = 0
\end{equation*}
Hence the columns of $\Kp$ determine \emph{pathways}: combinations of non
zero flows which lead to constant species amounts. In this example,
there is no $\Kp$ such that $N\Kp=0$. However, in the special case that
the four species $A$~-~$D$ are \emph{chemostats} (their values are
held constant by some external flows) then $N$ now has four \emph{zero}
entries and $\Kp=1$ and thus the four species have constant amounts for
any $v$ and thus the single reaction $r$ trivially becomes a
pathway. As discussed in the paper, the more complex systems of
Figures \ref{fig:HeiSch96}, \ref{fig:hill} \& \ref{fig:EskWriLoo05}
have non-trivial pathways.

Neither $G$ nor $\Kp$ are unique. The seminal contribution discussed by
\citet{HeiSch96} was to examine the case where the entries of $\Kp$ are
both \emph{integer} and \emph{non-negative} thus defining pathways
where all reaction flows have the same sign and are integer multiples
of some base flows. This idea forms the basis of analysing large-scale
biomolecular systems by \emph{flux-balance analysis} (FBA)
\citep{KliLieWie11}. Together with the bond graph concepts outlined in
\S~\ref{sec:short-intr-bond}, this idea  provides the foundation
for this paper.

\end{document}